# Effective Density of States Map of Undoped µc-Si:H Films: a Combined Experimental and Numerical Simulation Approach

Sanjay K. Ram[*] and Satyendra Kumar[!]
*Department of Physics, Indian Institute of Technology Kanpur, Kanpur-208016, India*

The phototransport properties of plasma deposited highly crystalline undoped hydrogenated microcrystalline silicon (µc-Si:H) films were studied by measuring the steady state photoconductivity (SSPC) as a function of temperature and light intensity. The films possessing different thicknesses and microstructures had been well characterized by various microstructural probes. µc-Si:H films possessing dissimilar microstructural attributes were found to exhibit different phototransport behaviors. We have employed numerical modeling of SSPC to corroborate and further elucidate the experimental results. Our study indicates that the different phototransport behaviors are linked to different features of the proposed density of states maps of the material which are different for µc-Si:H films having different types of microstructure.

PACS numbers: 73.50.Pz, 73.50.–h, 73.61.–r, 71.23.–k, 73.50.Gr, 72.20.Jv, 81.40.Gh

## I. INTRODUCTION

Plasma deposited hydrogenated microcrystalline silicon (µc-Si:H) has immense potential in large area electronic applications, as it offers the possibilities of high carrier mobilities[1] and better stability against light and current induced degradation as compared with amorphous silicon (a-Si:H).[2,3,4,5,6,7] These features, along with an ease of large area processing capabilities[2,8,9,10] even at low temperatures[11] make it an attractive candidate for use in solar cells[12] and thin film transistors.[13]

Microcrystalline silicon material is heterogeneous in nature consisting of crystalline and amorphous phases with presence of density deficit regions.[14,15] The microstructure of µc-Si:H is not unique due to the processing history, therefore, a comparison between various electronic transport models is futile, and there is still a long way to go in explaining the transport properties in the light of film microstructure. The presence of significant disorder in terms of variations in size and shape of crystallites (grains)[16] and nature of disordered phase (intergrain and inter-columnar boundaries)[17] complicates a comprehensive description of the optoelectronic properties in this material.[18,19,20,21] It is evident that smaller grain size imparts properties similar to those of a-Si:H, while large grain sizes confers properties closer to crystalline silicon.[18]

Thus presently, little is known about the recombination mechanisms and the nature / distribution of the density of gap states (DOS), and due to the above mentioned complexities, it is unreasonable to search for a unique *effective* DOS profile that would satisfy whole range of µc-Si:H materials and explain all the intricacies involved in its transport mechanisms.[22,23] Further, it needs to be emphasized that by analyzing a few samples of µc-Si:H produced under a narrow deposition regime, one cannot construct a DOS profile applicable to the whole class of µc-Si:H materials. Instead, it would be more scientific to construct DOS profiles based on film microstructure that would be applicable to a wider range of samples having some common unique microstructural features.[24,25,26]

The effects of the density and nature of gap states in semiconductors and insulators are extensively studied using photoconductivity and its recombination kinetics.[27,28] Transient photoconductivity (TPC)[29,30] and drift mobility measurement techniques[31,32,33,34] have been used to probe exponential conduction band and valence band tail (CBT and VBT respectively) states in µc-Si:H films. In addition, subgap absorption spectroscopy techniques such as constant photocurrent method (CPM),[33,35,36,37,38] dual beam photoconductivity (DBP),[39] and photothermal deflection spectroscopy (PDS)[37,38,40] are used to characterize the VBT states. Electron spin resonance (ESR) has also been used to probe the midgap states in µc-Si:H.[41] Some attempts have been made to decipher partial DOS distributions in the vicinity of CB edge using these probes.[5,42,43,44,45]

In contrast to the above methods that study only a specific portion of the DOS in the gap, steady state photoconductivity (SSPC) is an efficient and easy technique to examine the gap states over a wider range in the bandgap, and is sensitive to both density and nature of all the defect states acting as recombination centers between the quasi Fermi levels in the band gap. SSPC allows a comprehensive study of the phototransport properties of the electronic system of the corresponding material. The observed effects with this method can be the consequences of several different phototransport processes occurring in the system. Therefore, it is difficult to attribute the observed behavior to a particular effect. Certain fea-tures of SSPC in *a*-Si:H can be understood using the famous Rose model,[27] but this model is not always applicable in μc-Si:H films.[25]

[*] Corresponding author. E-mail address: skram@iitk.ac.in; sanjayk.ram@gmail.com
[!] satyen@iitk.ac.in

Numerical modeling has been extensively used to elicit information about the recombination kinetics and explain the experimental photoconductivity results in *a*-Si:H[46,47,48,49,50,51] and has added greatly to our knowledge of the nature and density of gap states in *a*-Si:H. In contrast, numerical modeling has been less employed to understand the SSPC in heterogeneous μc-Si:H.[22,24,25] However, efforts have been made to numerically model the transient photoconductivity (TPC),[29] time-of-flight (TOF) measurement[52,53] and modulated photoconductivity (MPC) processes[43] in μc-Si:H system. As discussed above, the photoelectronic properties of μc-Si:H are still inadequately understood due to a lack of knowledge about the DOS maps of μc-Si:H. In 2004, Balberg *et al.*[22] proposed the DOS of a single-phase μc-Si:H material using a comprehensive study involving SSPC experiments and numerical modeling that emphasized the dissimilarity of μc-Si:H with both polycrystalline and amorphous silicon.

In this paper, we report the findings of our study of phototransport properties of microstructurally different μc-Si:H films employing both experimental methods and numerical modeling of SSPC, and propose the effective DOS maps of these materials. For this purpose, we have first structurally characterized the μc-Si:H films using a variety of tools to elucidate a comprehensive picture of film microstructure and morphology. The dark electrical transport properties of these films, correlative with the microstructural findings led us to classify the material broadly into three types, having distinct microstructural and morphological attributes, with a particular dark electrical transport behavior peculiar to each class. The results of the structural and dark conductivity studies have been reported elsewhere,[54,55,56] but the findings have been summarized in §III.A, as they lay the foundation for the correlation between the structural aspects and phototransport properties.

The phototransport properties of these well-characterized films were then studied using SSPC and CPM techniques, the findings of the latter are not reported in this paper. The SSPC data was analyzed qualitatively, with a view to correlate the observed phototransport properties of each type of μc-Si:H material to its microstructural findings. However, elucidation of some other important phototransport properties such as recombination traffic, the role of different gap states in recombination process, and the complete DOS distribution in these microstructurally different μc-Si:H films require a numerical modeling study of the phototransport properties. In order to explore these aspects that would shed light on the phototransport properties of μc-Si:H and also substantiate the experimentally elucidated facts, we conducted a numerical modeling study of these three types of μc-Si:H.

This paper is organized as follows. In Sec. II, the theory of photoconductivity is presented. The Sec. III describes the SSPC experimental details, results and qualitative analysis. Numerical modeling study is described in Sec. IV, including the background, basic formalism, simulation procedure, and data analysis. Finally, the overall findings of the qualitative and quantitative analyses are summed up in Sec. V.

## II. THEORY

Photoconductivity can be described to consist of three mechanisms: First, absorption of photons and generation of free electron-hole pairs; second, transport of mobile carriers; and third, recombination of excess free electrons and holes through recombination centers.[27,28] Photoconductivity under steady state illumination is given by:

$$\sigma_{ph} = e[\mu_n(n - n_0) + \mu_p(p - p_0)], \quad (1)$$

where $\mu_n$ and $\mu_p$ are free-electron and hole mobilities respectively; $n$ and $p$ represent the steady state concentration of photoexcited electrons and holes, whereas $n_0$ and $p_0$ are their corresponding concentrations in the dark at a given temperature $T$. Since undoped μc-Si:H shows *n*-type behavior and $\mu_n > \mu_p$, electrons are taken to be the majority carriers.

In general, photoconductivity exhibits a non-integer power law dependence on carrier generation rate $G_L$ over several orders of magnitude given by:

$$\sigma_{ph} \propto G_L^{\gamma}, \quad (2)$$

The generation rate $G_L$ is determined by the external parameters and internal material parameters as:

$$G_L = \frac{\phi(1-R)}{d}[1 - e^{-\alpha d}] \quad (3)$$

where $\phi$ is the flux of photons in cm$^{-2}$s$^{-1}$, $R$ is the reflection coefficient at $h\nu = E$, $\alpha$ is the absorption coefficient of the material at energy $h\nu = E$, and $d$ is the film thickness.

The photoconductivity light intensity exponent, $\gamma$ provides information about the recombination mechanisms in a semiconductor material. $\gamma = 0.5$ represents bimolecular recombination kinetics, where electrons in the CB directly recombine with the holes in the VB. $\gamma = 1$ represents monomolecular recombination, that is, electrons in the CB recombine with holes in the VB, through the recombination centers in the gap. For $\gamma$ value lying between 0.5 and 1, Rose[27] described a model in which the density of trapped electrons $n_t$, approximated by

$$n_t(T) = \int_{-\infty}^{E_{fn}(T)} g(E)dE, \quad (4)$$

tracks the density of positively charged recombination centers $P_r$ to maintain charge neutrality. Rose showed that if the discrete states are distributed exponentially in the vicinity of band edges in the form of $e^{-\Delta E/kT_C}$, the photocurrent and light intensity curve should have the power $\gamma$ as:

$$\gamma = \frac{kT_c}{(kT + kT_c)}, \tag{5}$$

here $\Delta E$ is measured from the bottom of the CB and $kT_c$ is the characteristic energy of CBT *greater than or equal* to the energy corresponding to the measurement temperature $T$. In that case, most of the trapped electrons reside within $kT$ of the steady state Fermi level $E_{fn}$. The model as given by Eq. (4) is also applicable to the cases where DOS profile $g(E)$ does not decay in a purely exponential manner. The values for $\gamma$ then correspond to certain positions of $E_{fn}$, and an approximating exponential function with a local band tail parameter $kT_0(E_{fn})$ in its vicinity, giving $g(E_{fn})$.

The quasi Fermi level can be determined from the photoconductivity in the same range as was used to evaluate $\gamma$. The expression is given by:

$$[E_c - E_{fn}(\phi,T)] = [E_c - E_f(T)] - kT \ln\left[\frac{\sigma_{ph}(\phi,T)}{\sigma_d(T)}\right]$$

The above expression can be re-written in the way as below:

$$E_c - E_{fn} = kT \ln\left[\frac{\sigma_0}{\sigma_{ph}(T)}\right] \tag{6}$$

## III. EXPERIMENT: STEADY STATE PHOTOCONDUCTIVITY

### III.A. Experimental details

We prepared a series of highly crystallized undoped $\mu$c-Si:H films having varying degree of crystallinity by depositing on Corning 1737 substrates at a substrate temperature of 200°C in a parallel-plate glow discharge plasma deposition system operating at a standard rf frequency of 13.56 MHz using high purity $SiF_4$, Ar and $H_2$ as feed gases.[54,57] For the structural investigations, we employed variety of structural probes like in-situ spectroscopic ellipsometry (SE), Raman scattering (RS, from film and substrate side), X-ray diffraction (XRD) and atomic force microscopy (AFM). These well-characterized films were studied for the electron transport behavior using dark conductivity and photoconductivity as functions of several discerning parameters such as temperature, wavelength and intensity of probing light.[24,25] The effect of light intensity variation on the steady state photoconductivity was probed using above-bandgap light (He-Ne laser, $\lambda$ = 632.8 nm) in the temperature range of 20K–324K. Photon flux $\phi$ was varied from $\approx 10^{11}$ to $10^{17}$ photons/$cm^2$-sec using neutral density filters giving rise to generation rates of $\approx G_L = 10^{15}$-$10^{21}$ $cm^{-3}s^{-1}$. Penetration depth of this light is $\approx$500 nm. Many of the $\mu$c-Si:H films used in this study have been characterized by the time resolved microwave conductivity (TRMC) measurements.[58] TRMC is known to measure the mobility of carriers within the grains.[58]

### III.B. Findings of structural and dark electrical transport studies

High crystallinity of all the samples was confirmed by RS and SE measurements. SE data shows a crystalline volume fraction >90% from the initial stages of growth, with the rest being density deficit having no amorphous phase, and a reduced incubation layer thickness. RS results show a sharp peak at $\approx$522.6 $cm^{-1}$, corresponding to TO (transverse optical) mode in c-Si, without any amorphous phase that typically accompanies the c-Si peak in the RS profiles of $\mu$c-Si:H material. The detailed composition of the films educed from SE data shows grains of two distinct sizes, which is corroborated by the deconvolution of RS profiles using a crystallite size distribution of large grains (LG, $\approx$70-80 nm) and small grains (SG, $\leq$ 6-7 nm).[54,55] The bimodal crystallite size distribution is further supported by the XRD results showing large and small sized grains with different orientations.[59,60] There is significant variation in the percentage volume fraction of SG ($F_{cf}$) and LG ($F_{cl}$) with film growth. Preferential orientation in (400) and (220) directions is achieved by optimizing the deposition conditions leading to smooth top surfaces (surface roughness < 3 nm). Smooth top surfaces with less defect densities are highly attractive properties for device applications.[54,55]

Based on the structural investigations of the $\mu$c-Si:H films at various stages of growth, we were able to segregate out the unique features of microstructure and growth type present in the varieties of $\mu$c-Si:H films from the co-planar electrical transport point of view. All the $\mu$c-Si:H samples were classified into three types: *type-A*, *type-B*, and *type-C*, where we see the influence of the nature of inhomogeneities in $\mu$c-Si:H on the temperature dependent dark conductivity [$\sigma_d(T)$].[56] Our findings indicate that since deposition parameters have only an indirect causal link to the electrical properties through their primary effect on the microstructure of material, it is useful to consider the fractional composition of constituent large crystallite grains ($F_{cl}$) as a simple yet physically rational microstructural parameter that indicates the microstructural and morphological condition of the fully crystallized single phase $\mu$c-Si:H films and thus correlates acceptably with the electrical transport behavior.[55,56]

To summarize this classification, the *type-A* films have small grains, high density of inter-grain boundary regions containing disordered phase, and low amount of conglomeration. In this type, $F_{cl}$ <30%, $\sigma_0$ and $E_a$ are constant [$\approx 10^3$ ($\Omega cm$)$^{-1}$ and $\approx$0.55 eV respectively]. The *type-B* films contain a fixed ratio of mixed grains in the bulk. Conglomeration of grains results in a marked morphological variation, and a moderate amount of disordered phase in the conglomerate boundaries limits the electrical transport. Here $F_{cl}$ varies from 30% to 45%, there is a sharp drop in $\sigma_0$ [from $\approx 10^3$ to 0.1 $(\Omega cm)^{-1}$] and $E_a$ (from $\approx 0.55$ to 0.2 eV). The *type-C* μc-Si:H material is fully crystallized and crystallites are densely packed with significant fraction of large crystallites (>50%) and preferential orientation is seen. Here $\sigma_0$ shows a rising trend [from 0.05 to 1 $(\Omega cm)^{-1}$] and the fall in $E_a$ is slowed down (from 0.2 to 0.10 eV).[54,56]

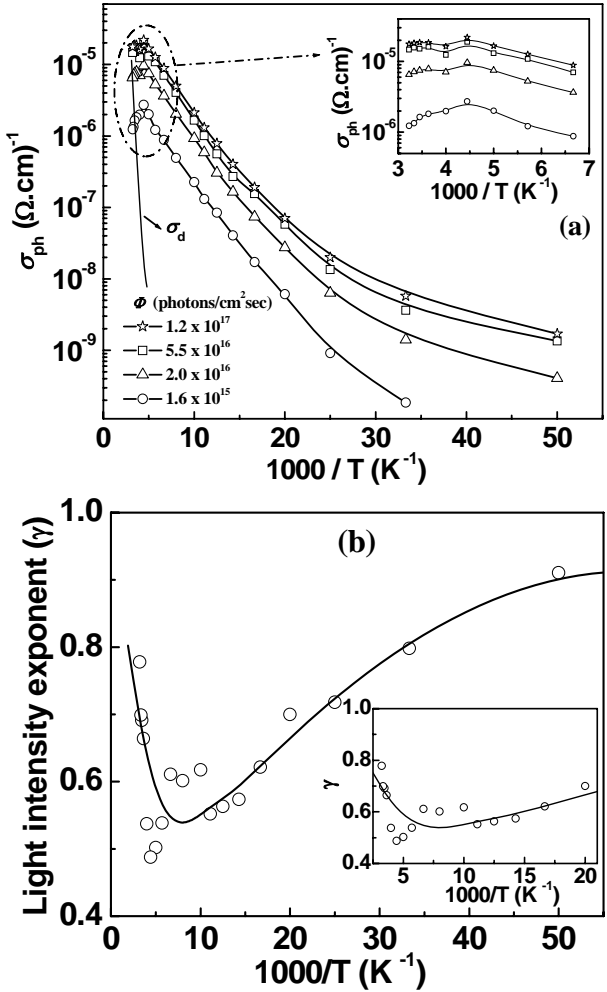

FIG. 1. (a) Temperature dependence of $\sigma_{ph}(T)$ of sample #B22 (*type-A*) for various light intensities ($\phi$) are shown by line+symbol. The temperature dependence of $\sigma_d$ of the same sample is shown by solid line. The inset shows the zoomed view of $\sigma_{ph}(T)$; (b) Temperature dependence of light intensity exponent $\gamma$ (obtained from light intensities dependence of $\sigma_{ph}(\phi)$ at different temperatures) of the same sample. The inset shows the zoomed view of $\gamma(T)$ from 325K down to 50K. Here the line is to guide the eye.

### III.C. Results of SSPC studies

Steady state photoconductivity and constant photocurrent method measurements were carried out on well-annealed samples using coplanar geometry in different experimental set-ups. To probe the effect of microstructural inhomogeneity on the phototransport properties of undoped μc-Si:H, we selected a few samples of each category of microstructure as described above. Though the studies were carried out on many samples, here we report the phototransport measurement results of three samples: sample #B22, sample #B23 and sample #F06, which are representative of *type-A*, *type-B* and *type-C* μc-Si:H materials respectively. The results of all the studied samples showed similar trends specific to the type of material. The details of calculation procedure to determine $E_f$ position in the gap[20,22,61] of these samples along with their electrical transport parameters are mentioned in the Table I.

#### Type-A μc-Si:H

The results of temperature dependent photoconductivity for various light intensities, $\sigma_{ph}(T,\phi)$ for sample #B22 (*type-A*) are shown in Fig. 1 (a). The temperature dependent dark conductivity of this sample is shown by a solid line in Fig. 1(a). The $\sigma_d(T)$ shows an activated behavior over a large temperature range ($\approx$170–450 K) with an activation energy $E_a$=0.5 eV. The $\sigma_{ph}(T)$ of this sample is essentially an increasing function of temperature for any value of light intensity used in the range mentioned above. However, at higher temperatures, $\sigma_{ph}$ is seen to decrease with increasing temperature, an effect known as thermal quenching (TQ). This effect is clearly seen in the inset of Fig. 1(a) over a temperature range of $T$ = 324–175K. The power law behavior of the photocurrent of this sample with a change in photon flux was observed throughout the temperature range of our study. The temperature dependence of light intensity exponent, $\gamma(T)$ calculated at each measurement temperature from the light intensity dependence of photoconductivity, $\sigma_{ph}(\phi)$ of the same sample is shown in Fig. 1(b). The inset shows the zoomed view of $\gamma(T)$. The variation of $\gamma$ is found to be between 0.5 and 1 in the whole temperature range.

#### Type-B μc-Si:H

The $\sigma_{ph}(T,\phi)$ for the sample #B23 (*type-B*) is shown in Fig. 2(a). The $\sigma_{ph}(T,\phi)$ behavior of #B23 is quite different from that of the above case as $\sigma_{ph}$ is found to increase monotonically with temperature without any thermal quenching effect. The solid line in this figure represents the dark conductivity with an activation energy $E_a$=0.34 eV. The $\gamma(T)$ values obtained for different temperatures for this sample are plotted against reciprocal of $T$ and depicted in Fig. 2(b). The $\gamma$ value of this sample is found to vary from 0.5 to 1 and it never goes down below the value 0.5 in the whole temperature range.

#### Type-C μc-Si:H

Fig. 3(a) shows the temperature dependence of $\sigma_{ph}$ for the sample #F06 (*type-C*) at various light intensities.

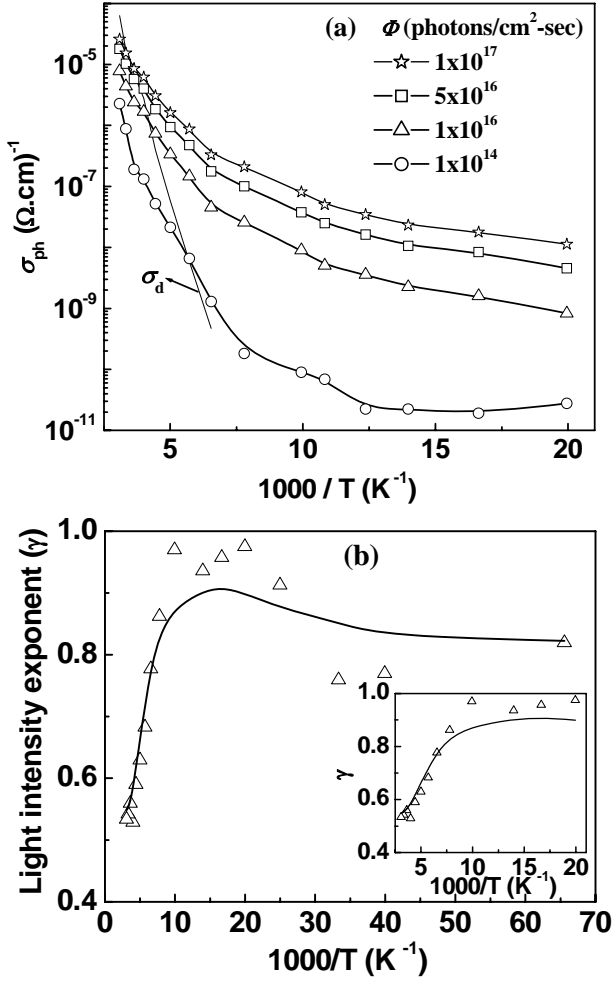

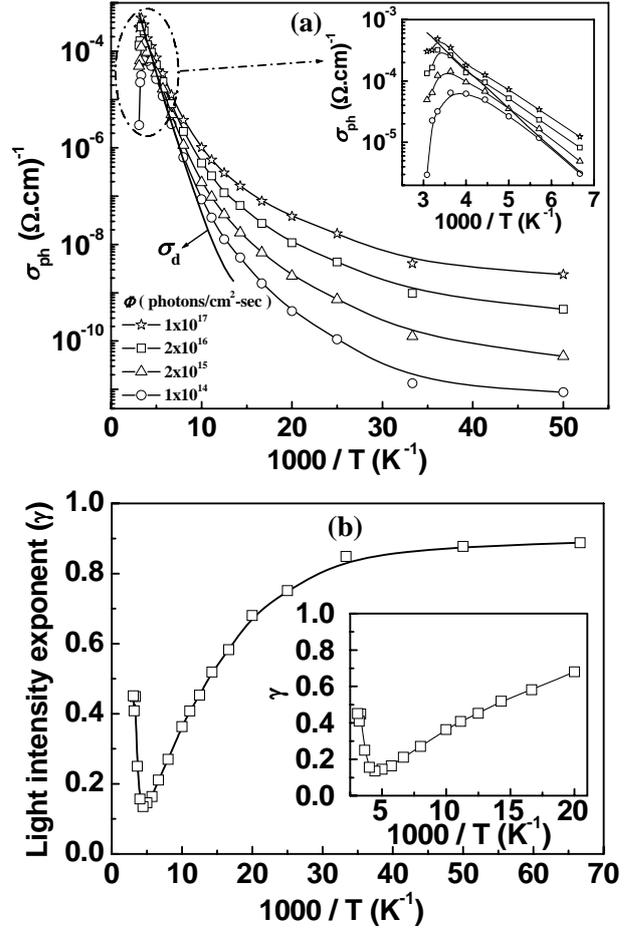

FIG. 2. (a) Temperature dependence of $\sigma_{ph}(T)$ of sample #B23 (*type-B*) for various light intensities are shown by line+symbol. The temperature dependence of $\sigma_d$ of the same sample is shown by solid line; (b) Temperature dependence of $\gamma(T)$ of the same sample. The inset shows the zoomed view of $\gamma(T)$ from 325K down to 50K. Here the line is to guide the eye.

FIG. 3. (a) Temperature dependence of $\sigma_{ph}(T)$ of sample #F06 (*type-C*) for various light intensities are shown by line+symbol. The temperature dependence of $\sigma_d$ of the same sample is shown by solid line. The inset shows the zoomed view of $\sigma_{ph}(T)$; (b) Temperature dependence of $\gamma(T)$ of the same sample. The inset shows the zoomed view of $\gamma(T)$ from 325K down to 50K.

The temperature dependent dark conductivity of the same sample is also shown by a solid line in the figure. The $\sigma_d(T)$ shows an activated behavior over a large temperature range (≈170 – 450 K) with an activation energy $E_a$=0.12 eV. The $\sigma_{ph}(T)$ of this sample monotonically increases with the increase in temperature excluding the lowest temperature region (<30K) where $\sigma_{ph}$ is nearly independent of *T*. However, at higher temperatures, thermal quenching effect is observed in the $\sigma_{ph}(T)$ behavior of this sample, similar to the kind of $\sigma_{ph}(T)$ behavior seen in sample #B22 (*type-A*). This effect is clearly seen for the temperature range of *T*=324 – 175K in the inset of Fig. 3(a). The maximum in $\sigma_{ph}(T)$ dependence shifts with increasing light intensity to higher temperatures where the dark current also increases, limiting the measurements at high temperatures. The $\gamma$ values obtained for different temperature are plotted in Fig. 3(b). While $\gamma$ never reaches 1 in the whole temperature range, we observe an unusually low value of $\gamma \approx 0.13$ at about 225 K. This anomalous behavior in photoconductivity data on undoped $\mu$c-Si:H has been observed in all of our fully crystallized samples representing *type-C* microstructure. It needs to be mentioned here that at higher light intensity ($\phi$>5×10$^{16}$ photons/cm$^2$s), $\gamma$ value further reduces to 0.05 for all the temperatures.

### III.D. Qualitative analysis of experimental SSPC data

It is interesting to note that the behavior of the phototransport properties of the samples belonging to the three types of $\mu$c-Si:H material are different. In brief, we can summarize the major outcome of the observed photo-

Table I. Details of calculation procedure to deduce $E_f$ position in the gap.

| Types, Sample# (R, thickness) | $\sigma_d$ $(\Omega.cm)^{-1}$ | $E_a$ (eV) | $\sigma_0$ $(\Omega.cm)^{-1}$ | $\mu$ $(cm^2/V\text{-}s)$ TRMC experiment | Reported method of $E_f$ calculation | |
|---|---|---|---|---|---|---|
| | | | | | Method-1 $E_c\text{-}E_f$ (eV) | Method-2 $E_c\text{-}E_f$ (eV) |
| Type-A #B22 (1/10, 170nm) | $3\times10^{-6}$ | 0.42 | 31.8 | 1.1 | 0.46 | 0.46 |
| Type-B #B23 (1/10, 590 nm) | $2\times10^{-5}$ | 0.3 | 1.71 | 4.4 | 0.42 | 0.41 |
| Type-C #F06 (1/1, 920nm) | $4.25\times10^{-4}$ | 0.12 | 0.08 | 6.2 | 0.34 | 0.33 |

**Method-1:** Considering the gap, $E_g = (E_c - E_v) = 1.8\,\text{eV}$

$E_c - E_f = E_a - kT\ln(\sigma_0/\sigma_o^s)$; where $\sigma_o^s$ is the standard value of dark conductivity prefactor $\sigma_0 = 150$ $(\Omega.cm)^{-1}$ for $\mu$c-Si:H system [Ref. 22]

**Method-2:** Considering the gap, $E_g = (E_c - E_v) = 1.12\,\text{eV}$;

$E_f - E_i = kT\ln(\sigma_d/\sigma_i)$; Standard value of intrinsic dark conductivity $\sigma_i = 6\times10^{-8}$ $(\Omega.cm)^{-1}$ and intrinsic energy $E_i - E_v = 0.56\,\text{eV}$ for $\mu$c-Si:H system [Ref. 20].

transport properties in the three types of samples in the following way:

a) In *type-A* material, light intensity exponent ($\gamma$) lies between 0.5 and 1, and temperature dependent photoconductivity [$\sigma_{ph}(T)$] shows thermal quenching (TQ) effect.
b) In *type-B* material, $0.5 < \gamma < 1$, with no TQ effect is observed.
c) In *type-C* material, $0.15 < \gamma < 1$, with a TQ effect is observed. This sublinear behavior of $\gamma$ and simultaneous presence of TQ effect is anomalous.

Therefore, it is now clear that no single phototransport mechanism or model can explain the results of the samples of all the three sets. However, it is highly possible that different phototransport behavior and mechanisms are taking place for different microstructures and thicknesses in such a heterogeneous system. Some of the above mentioned experimental findings of phototransport properties in $\mu$c-Si:H films, thermal quenching of photoconductivity and sublinear behavior of $\gamma$, are also observed in *a*-Si:H. To draw any valid parallel comparison between these two systems of materials, it would be helpful first to recall the well established models to understand the phototransport mechanism in a homogeneous system like *a*-Si:H, which we are summarizing here.

### *Photoconduction in a-Si:H*

The *TQ* effect in temperature dependent photoconductivity is observed in crystalline semiconductors[27] and in *a*-Si:H.[48,50,62] The phenomenon of *TQ* in crystalline photoconductors was explained by Rose using a model of two levels of states, one having larger capture coefficients for the majority carriers than the other level. The asymmetry in the capture coefficients causes *TQ* to occur in $\sigma_{ph}(T)$.[27] This theory is not applicable in case of amorphous semiconductors having continuous density of gap states. In *a*-Si:H these gap states consist of VBT, CBT and dangling bond (DB) defects. Over the past few decades, various models have been proposed to explain *TQ* in *a*-Si:H, which progressively improved as the knowledge of defect mechanisms, especially DBs, expanded.[47,49,63,64,65,66] In 1994, Tran[50] reported his extensive work to explain *TQ* in terms of the transference of recombination traffic from VBT to DBs. This work showed that *TQ* arises naturally from the asymmetry of band tails and the presence of DB states in the bandgap of *a*-Si:H. Tran's simulation results showed that *TQ* occurs even when all capture coefficients are identical, in contrast to the Rose model where they are larger in one of the two discrete levels. Usually *TQ* occurs for $T > 100$ K in *a*-Si:H. Tran expounded on the well-known experimental observation that the onset of *TQ* in *a*-Si:H shifts to higher temperature when defect densities are less or n-type doping level is increased.[67]

Let us now look at another important feature of the phototransport parameter, light intensity exponent $\gamma$. At low light excitation, unsaturation of recombination centers can lead to superlinearity while sublinear behavior can be observed at high light intensity when the saturation of such centers takes place. The explanation of sublinear photoconductive behavior is rather more controversial, and the model published include variations on the classic 'Rose model',[27] bimolecular recombination,[68] the shift of Fermi level $E_f$ towards band edges,[69,70] and the influence of surface defects and surface band bending.[71,72]

Another explanation of the sublinearity in $\sigma_{ph}(\phi)$ of doped a-Si:H, as proposed by Main et al.,[51] is that if CBT is steeper than VBT (i.e. $kT_c \ll kT_v$) and excess photocarriers are comparable to thermal carriers in dark, then Rose model does not hold; $\gamma$ at low excitation then becomes $T/T_v$ and at high excitation it changes to a constant value $T_c/T_v$.[51] Actually, with an increase in light intensity, the trapped hole quasi-Fermi level moves toward the band edge. To balance the resulting increase in trapped hole density, the trapped electron Fermi level moves upward, but by a smaller shift in energy, since the CBT is steeper than the valence band tail. Thus, the rate of increase of excess electrons is much smaller than the (linear) rate of increase of excess holes with light intensity. These deep hole traps are comparable to the 'safe hole traps' discussed by McMahon and Crandall.[73] The safe hole traps have lower capture coefficient for electrons compared to a much higher rate coefficient for electron capture to DBs. Recombination of holes trapped in such states involves, first an emission to the valence band or shallow valence band tail states, followed by capture by negatively charged DB state.[74]

### Photoconduction in µc-Si:H

Several types of phototransport behaviors have been observed in doped, undoped[22,41,44,75] or compensated µc-Si:H films.[76] In some cases, $TQ$ in the temperature dependent photoconductivity measurement has been observed at ≈ 250K whereas monotonic behavior of photoconductivity has also been reported.[22] Similarly, different behaviors of temperature dependence of light intensity exponent $\gamma$ have been found.[22,76] In addition, there are reports of different phototransport mechanisms operating in different temperature ranges.[77] The constant behavior of $\sigma_{ph}(T)$ and $\gamma \approx 1$ of µc-Si:H films in low temperature regime have been satisfactorily explained by energy loss hopping of photoexcited electrons and holes via localized band tail states before non-geminate recombination, similar to a-Si:H case. However, no such agreement is found among the reports that have attempted to explain the photoconductivity mechanisms in the medium and high temperature regime.

### III.D.1. Photoconductivity exponent: applicability of Rose model

The information about the distribution of gap states near the band edges of majority carrier is drawn from the exponent $\gamma$. According to Rose model,[27] when the $\gamma$ value lies between 0.5 and 1, it is assumed that the discrete states are distributed exponentially in the gap. The model is also found to be valid in a-Si:H where the localized states are continuous in the gap and it has been successfully used to explain several phototransport mechanisms in the a-Si:H material.[78] Since amorphous or disordered phase present in the boundary regions is one of the constituents of the complex microstructure of µc-Si:H films, it is likely to influence the phototransport behavior of these films as well. Bruggemann[44] has applied Rose model successfully in µc-Si:H films and obtained information about the localized states in the vicinity of CB edge and their effective DOS profile. Therefore, it is important to see if Rose model works for the varieties of highly crystallized µc-Si:H material used in our study.

To begin with, we try to apply the model to the results of the phototransport studies of the *type-A* material (sample #B22) shown in Fig. 1(a) and (b). The band-tail parameter $kT_c$ was calculated from the $\gamma$ values at different temperatures as shown in Fig. 1(b) by applying the Eq. (5). The Fig. 4(a) depicts the calculated $kT_c$. The quasi Fermi level was determined using Eq. (6), from those photoconductivity values that lie in the range of a particular constant value of $\gamma$ for any particular $T$. The condition, $kT_c > kT$ is maintained in the whole temperature range of our study. The temperature range for which the data was evaluated in terms of the model in Eq. (5)

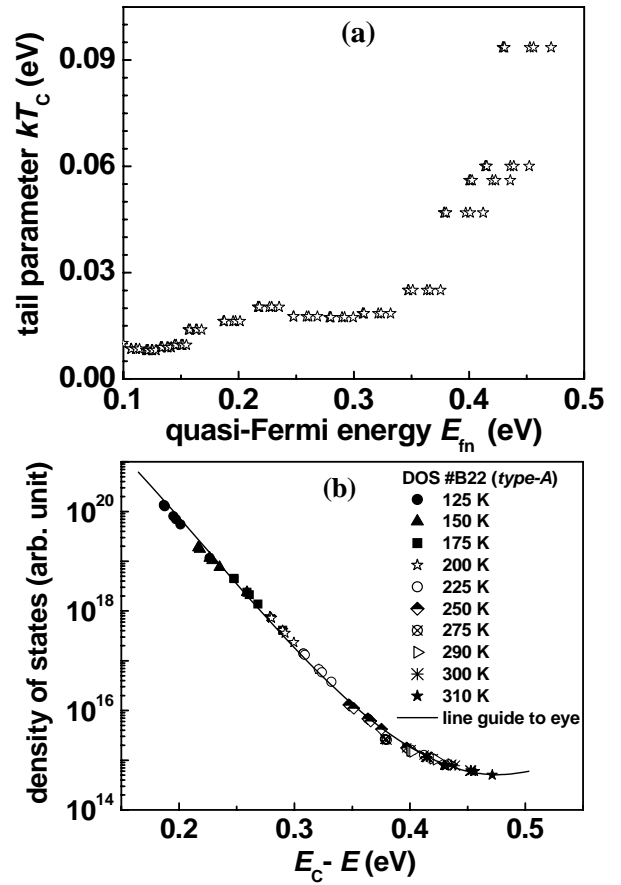

FIG. 4. (a) Plots of band-tail parameter $kT_c$ vs. quasi-Fermi energy $E_{fn}$ of sample #B22 (*type-A*), where $kT_c$ were calculated from the $\gamma$ values at different temperatures shown in Fig. 1(b); (b) Density of states (DOS) distribution obtained by fitting $kT_c$ vs $(E_c - E)$ data of part (a) to exponential distribution of states. This sketch of DOS profile in this sample is only an approximation.

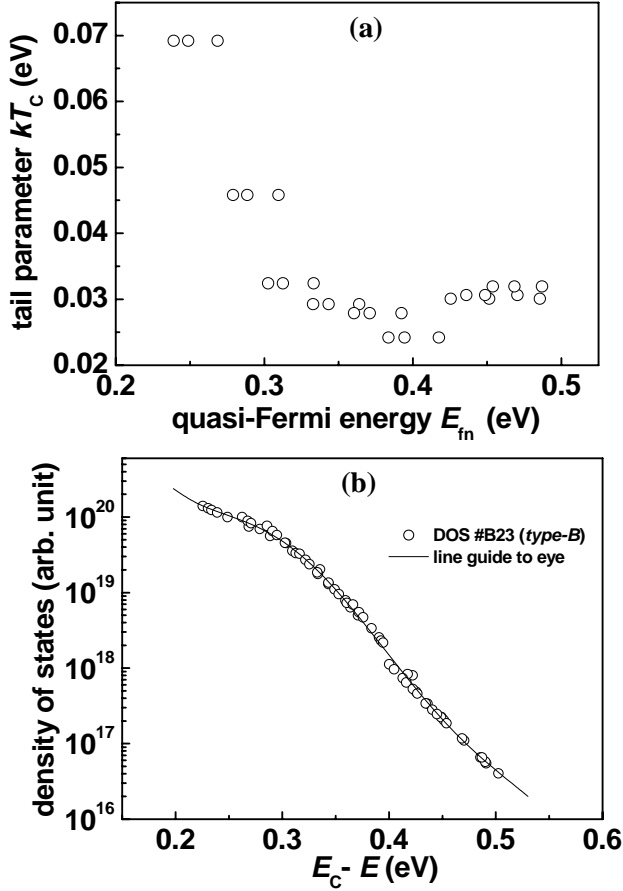

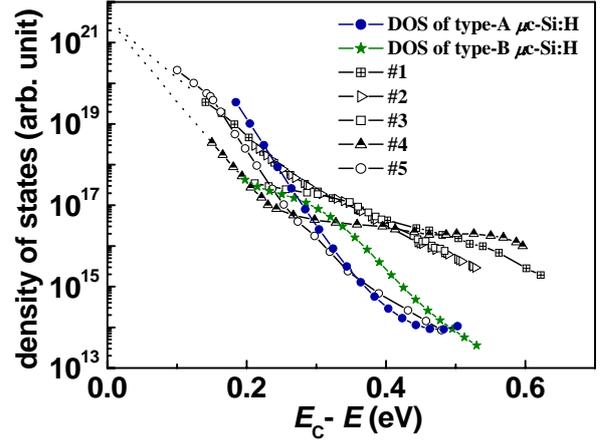

FIG. 5. (a) Plots of band-tail parameter $kT_c$ vs. quasi-Fermi energy $E_{fn}$ of sample #B23 (*type-B*), where $kT_c$ were calculated from the $\gamma$ values at different temperatures shown in Fig. 2(b); (b) Density of states (DOS) distribution obtained by fitting $kT_c$ vs $(E_c - E)$ data of part (a) to exponential distribution of states. This sketch of DOS profile in this sample is only an approximation.

FIG. 6. Density of states (DOS) distribution obtained for SSPC measurement of *types-A* and *B* µc-Si:H are plotted along with DOS profiles of µc-Si:H suggested in literature obtained with other experimental techniques (#1-5). This sketch of DOS profile in this sample is only an approximation. Here, #1 [Ref. 53; MPC-DOS coplanar µc-Si:H ($I_{crs}$=0.5)], #2 [Ref. 5,79; MPC-DOS HW-CVD µc-Si:H], #3 [Ref. 5,79; MPC-DOS SPC µc-Si:H], #4 [Ref. 52; TOF-DOS µc-Si:H], #5 [Ref. 44; SSPC-DOS µc-Si:H].

varies from 324 K down to 128 K. Lower values of $T$ were not used in the calculation as hopping conduction might be operating in that range.[54]

Finally, based on the values of band-tail parameter (at different temperatures and for different energetic positions) obtained from Fig. 4(a), we could roughly calculate the density of states below the CB edge for this *type-A* material. The approximated DOS profile is shown in Fig. 4(b). Here, the numerical values have been used only to embellish the characteristic energetic slopes with a visually appreciable structure of the DOS profile. It should be noted that this DOS profile is only a rough estimate, but it can help us visualize the shape of the localized band tail distribution near one of the band edges, in this case, the CB edge. The density of localized tail states is found to be exponentially distributed in the CB region, though at deeper energetic positions the DOS decays very slowly.

Now let us turn to the results of the phototransport studies of the *type-B* material (sample #B23) shown in Fig. 2(a) and (b). The method of calculation of band-tail parameter $kT_c$ and the quasi Fermi level at different temperatures was similar to the above case and the plot of values obtained is shown in Fig. 5(a). The condition $kT_c > kT$ is maintained in this case too. Here also low temperature region is not considered due to the possibility of hopping conduction operating in that range. The Fig. 5(b) shows a rough sketch of the density of states profile below the CB edge for the *type-B* µc-Si:H material. In this case, DOS is found to have slowly decaying states in the energy interval 0.2 to 0.3 eV, followed by a steeper tail. Similar to the DOS of *type-A*, the density of localized tail states at deeper energetic position shows less steep tail.

Therefore, we see that in samples of *type-A* and *B*, the behavior of $\gamma$ is in agreement with Rose model, suggesting the presence of a band-tail state distribution in these two materials. For comparison we have plotted our experimentally estimated DOS profiles of both the types of µc-Si:H along with the DOS profiles of µc-Si:H suggested in the literature derived from other experimental techniques (#1-4)[5,52,53,79] in Fig. 6. Though the obtained DOS profiles of our µc-Si:H samples are just rough sketches, but they are not unphysical and are in excellent agreement with the DOS obtained by other groups.[5,53,79] The shape of our DOS profile of *type-A* material is quite similar to the SSPC-DOS estimated by Bruggemann (#5).[44] However, the shape of DOS profile of *type-B* material near the band edge (from 0.2 to 0.3 eV) is similar to that of MPC-DOS obtained for solid phase crystallized (SPC) µc-Si:H (#3).[5,79] The reason could be due to the highly crystalline nature of our *type-B* material.

The lower values of DOS of our µc-Si:H at deeper energetic position in the gap can be the result of the passivation of grain boundaries by $H_2$, while the boundaries of SPC $\mu$c-Si:H material are not passivated, and can therefore give rise to high value of DOS at deeper energies in the gap. However, in *type-C* material, when we tried to fit Rose model to its phototransport results, viz., $\gamma$, we find a very narrow width of CBT ($kT_c \approx 0.01$ eV). Here, $kT_c$ is found to be less than $kT$ in the whole range of temperature, which is not a valid assumption for Rose theory. Therefore, one cannot apply this model to the *type-C* material. Now we shall discuss the results of each type of material in context of its characteristic microstructural features, dark conductivity properties and physically plausible DOS features that can give rise to such phototransport behavior.

### III.D.2. Phototransport behavior of Type-A μc-Si:H (TQ and 0.5<γ<1)

It can be recalled that *type-A* $\mu$c-Si:H films are crystallized with small grains. The smaller grain size results in a higher number density of boundary regions, the unsaturated DBs are assumed to be located in these disordered boundary regions. Therefore, the material may possess quite a significant number of DB densities, though less than what is present in *a*-Si:H. It is tempting to treat *type-A* $\mu$c-Si:H analogous to *a*-Si:H. However, there are certain striking differences in the phototransport mechanisms in the two materials. In particular, thermal quenching of $\sigma_{ph}(T)$ in an *undoped* a-Si:H is usually accompanied by superlinear behavior of $\gamma$.[48,50,62] In contrast, we see *TQ* accompanied by 0.5<$\gamma$<1 in case of *type-A* $\mu$c-Si:H.

Moreover, *TQ* in *a*-Si:H is seen near 100-150 K[50] whereas the onset of *TQ* in this case is near 225 K. As we have seen above, *TQ* effect in *a*-Si:H can occur because of the presence of DBs and the asymmetry in the band tail states. In fact, detailed simulation of $\sigma_{ph}$ by Tran indicated an increase in the onset temperature of *TQ* by reducing the DB density.[50] Regarding the asymmetry in the shape of both the band tail states, we find valuable information about at least one of the localized tail states as to its distribution in the vicinity of CB edge by using Rose model [Fig. 4(a)]. The results of the dependence of sub gap absorption on varying microstructure of $\mu$c-Si:H films suggest that the width of VBT of *type-A* material should be larger than *type-B* and *C* materials. The VBT width of *type-A* material is closer to that of *a*-Si:H. Therefore, we can conclude that the width of VBT is larger than that of CBT, giving rise to asymmetry in band tail states.

Now the problem remains to find the cause of absence of superlinear behavior of $\gamma$, which usually accompanies *TQ* in *a*-Si:H. The dark conductivity results of this sample[61] predict the Fermi level position to be around $E_f \approx (E_c - 0.46)$ eV. If we assume the energy gap of this disorder phase present in the boundary (where the recombination process is taking place) $\approx 1.8$ eV similar to the value in *a*-Si:H, then $E_f \approx 1.34$ eV above the VB edge (see Table I). In *a*-Si:H it has been observed that on shifting $E_f$ towards CB edge either by doping or any other means, the $\gamma$ value is found to decrease.[69,70] In this case $E_f$ is also at a higher position than found in undoped *a*-Si:H. This might be the reason why superlinear behavior was not observed. However, we can see a slight increase of $\gamma$ value near the temperature region of *TQ* onset. The rise in $\gamma$ value at high temperature is due to the effect of thermally generated carriers.

### III.D.3. Phototransport behavior of Type-C μc-Si:H (TQ and γ<0.5)

Observation of $\gamma$<0.5 along with a thermal quenching is unusual.[24,25] Though it has been observed in $\mu$c-Si:H,[80] but the simultaneous occurrence of the two phenomena has not been explored. Here we need to search for possible reasons that can satisfactorily explain occurrence of *TQ* with $\gamma$<0.5. *Type-C* $\mu$c-Si:H material consists of tightly packed large crystallites with preferred orientation and no trace of amorphous Si tissue. It is therefore difficult to link the observed *TQ* to just the DB density. Now, let us look at the role of asymmetry in band tails in the causation of *TQ*. For *type-C* $\mu$c-Si:H, we found $kT_c < kT$ (§ III.D.1). Therefore, a narrow width of CBT ($kT_c < 0.02$ eV) can be expected. According to the defect pool model, *n*-type doping in *a*-Si:H causes a large increase in negatively charged DBs density together with a decrease in positively charged dangling band states in the gap, which in turn shows a lower DOS near the CB edge leading to steeper CBT.[81,82,83] In *type-C* $\mu$c-Si:H, higher density of available free carriers and low value of defect density can also create a possibility for steeper CBT. Now let us look into the VBT region. According to reports, it has been found using constant photocurrent measurement that Urbach energy of the hole band tail increases with increasing crystallinity.[37] This results in a higher number of density of states in the deeper side of VBT causing the overall width of VBT to be larger than that of CBT. So it is evident that the slope of CBT must be steeper compared to VBT and hence can give arise to asymmetry in both the tail states, causing *TQ* to occur.

Another peculiarity observed is the sublinear behavior of $\gamma$ (<0.5) in *type-C* $\mu$c-Si:H for most of the temperature region except at very low temperature where $\gamma$ increases above 0.5 [Fig. 3(c)]. The sublinear behavior of $\gamma$ found in doped *a*-Si:H at higher excitation of light has been attributed to the shift of Fermi level $E_f$ towards band edges.[70] It has been reported that in $\mu$c-Si:H the localized states are lower compared to *a*-Si:H, thus making it possible for the $E_f$ to shift across the gap distribution, and therefore, the photoconductivity behavior can be largely affected by the position of $E_f$.[84] In our case, for this particular sample $E_f$ is found to be very close to $E_c$ ($E_c - E_f \approx 0.34$ eV) as calculated from dark conductivity $E_a$ measurement (see Table I).[61] This material is not highly photosensitive, and excess photo-carriers in it are comparable to thermal carriers in dark. Here, CBT is expected to be steeper than VBT (i.e., $T_c \ll T_v$), and Rose model was not found to hold in this system (see § III.D.1).[51] All these observations are very reminiscent of the experimental data and model simulation of Main *et al.* regarding sublinear behavior in doped *a*-Si:H.[51] Possibly the same explanation holds true for the observed sublinear behavior of $\gamma$ in *type-C* material too, in which case the very low value of $\gamma$ at higher illumination where $\gamma$ acquires a constant value $\approx T_c/T_v$ is also well explained.

Fully crystallized *type-C* μc-Si:H should be compared with what is known as hydrogenated poly-silicon. If we look into the DOS distribution of polycrystalline Si,[85] we find that the VBT in such a material has two distinct parts having different slopes; one with a sharper slope near the edge and another with a less steep slope at deeper energy. A similar DOS distribution has also been proposed by Vanderhaghen *et al.*[31] to explain the phototransport properties in highly crystallized μc-Si:H films having grains joined together in the percolation regime. From electronic transport point of view, an excellent percolation pathway is expected in *type-C* μc-Si:H. Therefore, effective DOS in fully crystallized μc-Si:H films of *type-C* may exhibit two different valence band tails; a sharper, shallow tail originating from grain boundary defects, and another less steep, deep tail associated with the defects in the columnar boundary regions. Capture cross section for the deeper VBT states is expected to be smaller than the shallower states.[31] This deeper tail can also work as "safe hole traps".[51] These trap centers are also reported to be the reason for the sublinear behavior of $\gamma$, as has been previously described in the context of *a*-Si:H.[22] This DOS profile is compatible with the explanations of sublinearity and *TQ* as mentioned above.

A detailed numerical simulation that is necessary to support the qualitative arguments proffered here is presented in coming section.

### III.D.4. Phototransport behavior of Type-B μc-Si:H (No TQ and 0.5<γ<1)

The phototransport properties observed in this case have typically been seen in μc-Si:H. According to Balberg *et al.*, semi-Gaussian distribution of VBT along with a narrow width is responsible for such observations of phototransport properties in undoped microcrystalline silicon.[22] Further, the absence of TQ emphasizes that both the tail states should be somewhat symmetric to each other. Regarding the shape of CBT in this type of material, the characteristic width was deduced to be ≈ 25-28 meV by applying Rose model (§ III.D.1), as was also observed in Ref. [44]. A rough sketch of the effective DOS profile in the vicinity of CB edge suggests that instead of a single exponential band tail of CBT, some hybrid form of DOS is approximated [Fig. 5(b)].

Let us now consider the possible shapes of DOS near VBT. The microstructural investigations of this type of material led us to envisage the films to be having mixed grains (both small and large) and moderate disorder phase in the columnar boundary regions. In the films of *type-B* μc-Si:H material, percolation paths are restricted. According to Vanderhaghen *et al.*,[31] the μc-Si:H films having unpercolated columns can also have two valence band tail slopes configuration like the films having higher number of percolated columns. Nevertheless, the value of the density of states of the deeper VBT will be lower in the film with unpercolated grains than in a film with percolated columns. In other words, the steeper shallow tail states in unpercolated columns plunge deeper as they fall from the band edge when compared to the case of percolated ones. After the initial fall, it associates with a deeper tail state arising from columnar boundaries.

To briefly recapitulate the experimental results presented and discussed so far, we have qualitatively explained all the phototransport findings in the context of the different microstructures of the three types of material and their possible DOS distributions, taking into account the various possibilities for the origin of the different phototransport behavior and the recombination mechanisms that can explain the findings comprehensively and plausibly. The results of phototransport studies of our μc-Si:H films presented in this section assert that in such a heterogeneous system, films of the same material having different microstructures and thicknesses can have different phototransport behavior, different underlying mechanisms and different effective DOS distributions. However, elucidation of some other important phototransport properties such as recombination traffic, the role of different gap states in recombination process, and the complete DOS distributions in these microstructurally different μc-Si:H films requires a numerical modeling study of the phototransport properties, which we have explored in the following section.

### IV. NUMERICAL MODELING OF SSPC IN μc-Si:H SYSTEM

A proper treatise on phototransport behavior in μc-Si:H is incomplete without a modeling study of photoconductivity, that would bring to light many aspects of the mechanisms involved. With this aim, we first constructed possible effective DOS distributions of these samples of different types by considering the qualitative arguments and analysis for experimentally observed phototransport properties for different microstructures as described in the previous section, and then carried out rigorous numerical simulations with sensitivity analysis using Shockley-Read statistics in steady state conditions to determine the recombination process.[86]

In this section, we first review the state-of-art modeling approaches to understand the phototransport properties of *a*-Si:H and μc-Si:H material in § IV.A. In the next section § IV.B, we have presented the basic formalism of our simulation model based on the recombination mechanisms in *a*-Si:H material. To check the program flow and correctness of our simulation procedure, we tested our simulation code on a model of Tran (model –B1; Ref. 50) and used the parameters and DOS profiles as mentioned in his work. Our results (not shown here) showed excellent reproducibility of the numerical values of carrier densities as obtained by Tran. It may be pointed out that unlike Tran, we have not made any assumptions and approximations (e.g., emission from gap states being negligible at low temperatures) to simplify the calculation procedure. This validation of our simulation procedure and methodology led us to proceed to apply this methodology in numerical modeling of phototransport mechanisms in our *µ*c-Si:H material. We applied the same procedure with different and suitable DOS profiles and parameters as appropriate for the experimentally observed phototransport behaviors of the microstructurally different *µ*c-Si:H films used in this work (§IV.C). Results of the numerical modeling are presented and analyzed in §IV.D.

### IV.A. Background

Shockley and Read described the statistics of a single trapping level in terms of four simple generation-recombination processes.[86] The approach is extremely successful in describing non-equilibrium steady-state processes in crystalline semiconductors. Later on, the recombination statistics for an arbitrary distribution of gap states under steady-state illumination condition was introduced by Rose[27] in a phenomenological approach. Further, Simmons and Taylor extended Shockley-Read statistics to more than one distinct trapping level and showed that the probability of occupation of a trap level at any energy is just the same as derived by Shockley-Read for a single trap level.[87,88] An arbitrary distribution of gap states can also be considered as an ensemble of single-level states and should therefore obey the same statistics for discrete level states at the same energy. However, it should be noted that this partition function is independent of the energy distribution of the traps in the gap.

The recombination statistics lay the foundation for our understanding of various phenomena and play a critical role in modeling the photo-transport behavior in disordered semiconductors. Numerical simulation studies of phototransport properties in amorphous silicon were mostly carried out to understand the thermal quenching (TQ) behavior and a variety of photoconductivity exponent ($\gamma$) values. Hack *et al.*[70] considered four exponential distributions in the gap states and explained the dependence of $\gamma$ on Fermi level. Later, Vaillant and Jousse explained this relationship by making a recombination model for simulation by including the statistics for correlated defects especially the DBs, and Shockley-Read and Simmons-Taylor statistics for the exponential band tails.[46] They considered the recombination takes place only in DB's and the exponential band tails behave as trap states. In 1988, Vaillant *et al.*[47] further added that recombination occurs at both DBs and band tails. Since then various workers have adopted this basic formalism in simulation models to understand a variety of phototransport phenomena.[89,90,91,92] In 1995, M.Q. Tran[50] presented a comprehensive and detailed recombination model to explain the thermal quenching behavior of photoconductivity process in *a*-Si:H. The role of various parameters involved in model simulation process and their effect on various phototransport properties were thoroughly addressed. Based on this work, Balberg *et al.*[22] carried out similar numerical calculations to explain the experimentally observed phototransport properties of *µ*c-Si:H.

### IV.B. Basic formalism of simulation model

#### IV.B.1. General assumption

We consider photon energy greater than the band gap and suppose that the photo-generated electrons and holes are excited above the mobility edges. Under illumination, tail states and DBs act as recombination centers and we assume that there are no other localized states in the gap active for recombination. We show a sketch of the possible transitions under illumination in Fig. 7. The details of each transition follow later in this section.

In the range of temperatures considered here, photoconduction is carried out by free carriers. So, only transitions between a recombination center and the bands are taken into account. Because of their small probabilities, neither the band-to-band transitions and recombination, nor the transitions between recombination centers have been considered. In addition, DBs are assumed to capture only mobile free carriers and because of similar reasons of small probabilities associated with DBs, the transition between two DBs is also not considered here.

#### IV.B.2. Principle of calculation

Photoconductivity under steady state illumination as described in Eq. (1), §II, is given by:

$$\sigma_{ph} = e\{\mu_n(n-n_0) + \mu_p(p-p_0)\},$$

where $\mu_n$ and $\mu_p$ are the free-electron and hole mobilities, respectively; $n$ and $p$ represent the steady state concentration of photoexcited electrons and holes whereas $n_0$ and $p_0$ are their corresponding concentrations in the dark at a given temperature $T$ and can be given as below:

$$n_0 = N_c \exp\left[-\frac{(E_c - E_f)}{kT}\right];$$

$$p_0 = N_v \exp\left[-\frac{(E_f - E_v)}{kT}\right] \quad (7)$$

If we assume the recombination of excess carriers

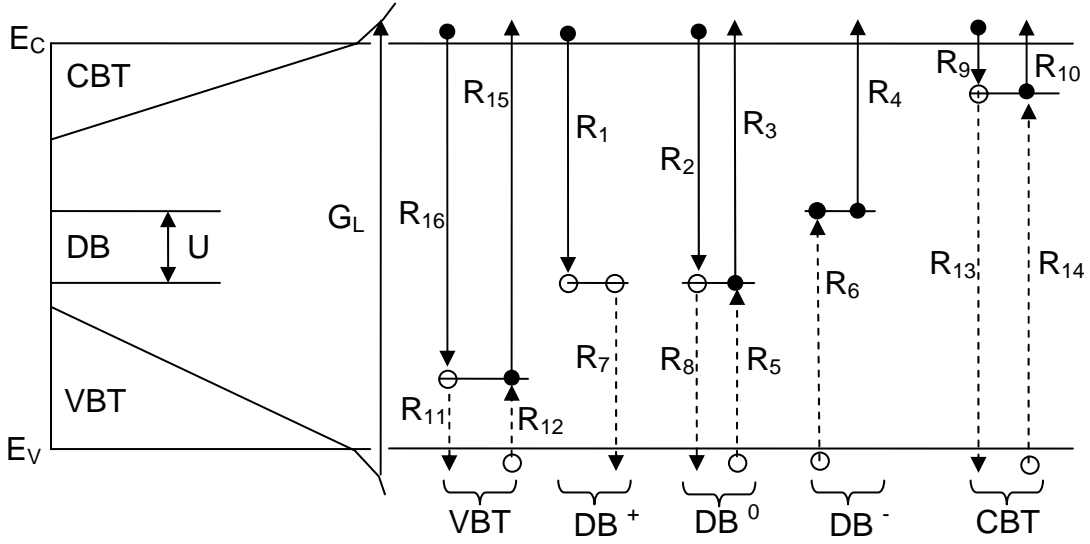

FIG. 7. Schematics of different recombination processes taking place within the gap of a disordered material.

takes place in CBT, VBT and in DBs, then the charge neutrality and the recombination rate equation can be expressed as below:

*The Charge Neutrality equation:*

$$n - p + Q_{CT}(n,p) - Q_{VT}(n,p) + Q_{DB}(n,p) = \\ n_0 - p_0 + Q_{CT}(n_0,p_0) - Q_{VT}(n_0,p_0) + Q_{DB}(n_0,p_0) \quad (8)$$

*The Recombination Rate equation:*

$$G_L = U_{CT} + U_{VT} + U_{DB} \quad (9)$$

In Eq. (8) $Q_{CT}(n,p)$, $Q_{VT}(n,p)$ and $Q_{DB}(n,p)$ are the carrier concentrations in the CBT, VBT and in DB respectively under illumination. Their respective thermal equilibrium values are $Q_{CT}(n_0,p_0)$, $Q_{VT}(n_0,p_0)$ and $Q_{DB}(n_0,p_0)$. The net recombination rates through the CBT, VBT and DBs are denoted by $U_{CT}$, $U_{VT}$ and $U_{DB}$ in Eq. (9). These are the rates of capture into the localized states less the rates of thermal emission from these states. Under steady state illumination, the densities $n$ and $p$ must simultaneously satisfy the above two conditions, namely that the total net charge in the sample is zero [Eq. (8)], and the total recombination rate equals the generation rate $G_L$ [Eq. (9)]. The details of these equations are given in the sections that follow.

### IV.B.3. Emission & capture process, partition functions, and rate equations

The flows of carriers through tail states and DB levels can be seen in Fig. 7. Here, the DOS includes two exponential tail states (CBT and VBT) arising from the disorder of the continuous random network and DB states situated around midgap. There is considerable experimental evidence that midgap defects (dangling bonds) in amorphous silicon material are amphoteric, exhibiting three possible charge states: A positively charged donor like state $DB^+$ with an energy $E_{DB}$, a neutral state $DB^0$ and a negatively charged acceptor like state $DB^-$ that lies at $E_{DB} + U$, where $U$ is the correlation energy. On the other hand, tail states behave as either donor or acceptor like states.

The carrier transitions in *DB* states are represented in the following manner:

$$DB^+ + DB^- \to 2DB^0$$

The transition of this type has very small probabilities and therefore has been neglected. However, the allowed transitions are between *DBs* and *VBT* or *CBT* given by:

$$DB^+ + e \leftrightarrow DB^0,$$

$$DB^0 + e \leftrightarrow DB^-,$$

*Under Thermodynamic Equilibrium Conditions:*

In the dark, at thermal equilibrium, the independent equations of conservation characterize the system represented in Fig. 7:

$$\frac{dn}{dt} = \overbrace{R_1 - R_3 + R_2 - R_4}^{DB} + \overbrace{R_9 - R_{10}}^{CBT} + \overbrace{R_{16} - R_{15}}^{VBT} = 0 \quad (10\text{-a})$$

$$\frac{dp}{dt} = \overbrace{R_7 - R_5 + R_8 - R_6}^{DB} + \overbrace{R_{14} - R_{13}}^{CBT} + \overbrace{R_{12} - R_{11}}^{VBT} = 0 \quad (10\text{-b})$$

$$\frac{d[DB^+]}{dt} = R_1 + R_7 - R_3 - R_5 = 0 \quad (10\text{-c})$$

$$\frac{d[DB^-]}{dt} = R_2 + R_8 - R_4 - R_6 = 0 \qquad (10\text{-d})$$

The different flows (thermal transitions per unit volume and unit time) are named as $R_i$ and will be expressed as a function of the emission and capture parameters extending the Shockley–Read expressions to amphoteric-like states:

$$R_1 = n N_{DB} F_{DB}^+ S_n^+ ; \; R_2 = n N_{DB} F_{DB}^0 S_n^0 ;$$
$$R_3 = N_{DB} F_{DB}^0 \varepsilon_n^0 ; \; R_4 = N_{DB} F_{DB}^- \varepsilon_n^- ;$$
$$R_5 = p N_{DB} F_{DB}^0 S_p^0 ; \; R_6 = p N_{DB} F_{DB}^- S_p^- ;$$
$$R_7 = N_{DB} F_{DB}^+ \varepsilon_p^+ ; \; R_8 = N_{DB} F_{DB}^0 \varepsilon_p^0, \qquad (11)$$

where $N_{DB}$ is the density of DBs in the material.

In these equations the following notations are used: $F_{DB}^+$, $F_{DB}^0$, and $F_{DB}^-$ are the occupation rates of DB's in, respectively, the $DB^+$, $DB^0$, and $DB^-$ states and $F_{DB}^{0+}$, $F_{DB}^{00}$, and $F_{DB}^{0-}$ are the same occupation rates at thermal equilibrium; $S_n^+$ and $S_n^0$ are the rate coefficients (cm$^3$s$^{-1}$) for electron capture by $DB^+$ and $DB^0$ states, respectively; $S_p^-$ and $S_p^0$ are the rate coefficients (cm$^3$s$^{-1}$) for hole capture by $DB^-$ and $DB^0$ states, respectively; $\varepsilon_n^-$ and $\varepsilon_n^0$ are the rates (s$^{-1}$) for electron emission by $DB^+$ and $DB^0$ states, respectively; $\varepsilon_p^+$ and $\varepsilon_p^0$ are the rates (s$^{-1}$) for hole emission by $DB^-$ and $DB^0$ states, respectively; The rates $S_n^+$, $S_n^0$, $S_p^-$, and $S_p^0$ are given by the well known relations $S_n^+ = v_{th}\sigma_n^+$, $S_n^0 = v_{th}\sigma_n^0$, $S_p^- = v_{th}\sigma_p^-$, and $S_p^0 = v_{th}\sigma_p^0$, where $\sigma_n^+$, $\sigma_n^0$, $\sigma_p^-$, and $\sigma_p^0$ are the corresponding capture cross sections and $v_{th}$ is the free carrier thermal velocity.

Similarly, $S_n^{CT}$ and $S_n^{VT}$ denotes the capture coefficient of free electros into the CBT and VBT respectively; and $S_p^{CT}$ and $S_p^{VT}$ are the capture coefficient of free holes into the CBT and VBT respectively;

The electron emission rate from CBT state is given by:

$$\varepsilon_n(E) = S_n^{CT} N_c \exp\left[-\frac{(E_c - E)}{kT}\right] \qquad (12)$$

where $N_c = N_{CT}^0 kT$ is the effective density of states at $E_c$.

Similarly, the hole emission rate from CBT is given by:

$$\varepsilon_p(E) = S_p^{CT} N_v \exp\left[-\frac{(E - E_v)}{kT}\right] \qquad (13)$$

where $N_v = N_{VT}^0 kT$ being the effective density of states at $E_v$. Sometimes, the expressions shown in Eqs. (12) and (13) are also written in the form:

$$\varepsilon_n(E) = S_n^{CT} n' \quad \text{and} \quad \varepsilon_p(E) = S_p^{CT} p' \qquad (14)$$

respectively for Eqs. (12) and (13), where

$$n' = N_c \exp\left[-\frac{(E_c - E)}{kT}\right]$$

and

$$p' = N_v \exp\left[-\frac{(E - E_v)}{kT}\right].$$

To obtain the emission rates for the VBT state, one replaces the superscript CT in Eqs. (12 – 14) with VT.

Under thermodynamic equilibrium conditions, the principle of detailed balance[93] must be satisfied, thus $R_1 = R_3$, $R_2 = R_4$, $R_5 = R_7$, and $R_6 = R_8$. Taking into account Eq. (10-c) and Eq. (10-d), this leads to the following equalities:

$$\varepsilon_n^0 = n_0 \frac{F_{DB}^{0+}}{F_{DB}^{00}} S_n^+ ; \; \varepsilon_n^- = n_0 \frac{F_{DB}^{00}}{F_{DB}^{0-}} S_n^0 ;$$
$$\varepsilon_p^0 = p_0 \frac{F_{DB}^{0-}}{F_{DB}^{00}} S_p^- ; \; \varepsilon_p^+ = p_0 \frac{F_{DB}^{00}}{F_{DB}^{0+}} S_p^0 \qquad (15)$$

where the subscript "0" stands for thermodynamic equilibrium. Using the grand partition function of the system, the occupation probabilities: $F_{DB}^{0+}$, $F_{DB}^{00}$, and $F_{DB}^{0-}$ under thermal equilibrium can be easily determined.[94]

$$\left.\begin{array}{l} F_{DB}^{0+} = \dfrac{1}{1 + 2\exp\left[\dfrac{(E_f - E)}{kT}\right] + \exp\left[\dfrac{(2E_f - 2E - U)}{kT}\right]} \\[2ex] F_{DB}^{00} = 2\exp\left[\dfrac{(E_f - E)}{kT}\right] \cdot F_{DB}^{0+} \\[2ex] F_{DB}^{0-} = 1 - F_{DB}^{00} - F_{DB}^{0+} \end{array}\right\} \qquad (16)$$

States from VBT and CBT will also contribute to the thermal transitions and will obey Fermi-Dirac statistics in the dark condition.

*Under Non-equilibrium Illumination Conditions:*

Under nonequilibrium steady-state conditions, we make the usual assumption that the emission and capture rates remain unaltered with respect to equilibrium conditions. Equations 10(a) – 10(d) in the steady state out of equilibrium become

$$\frac{dn}{dt} = G_L - \left[\overbrace{R_1 - R_3 + R_2 - R_4}^{DB} + \overbrace{R_9 - R_{10}}^{CBT} + \overbrace{R_{16} - R_{15}}^{VBT}\right] = 0$$

(17-a)

$$\frac{dp}{dt}=G_{L}-\left[\overbrace{R_{7}-R_{5}+R_{8}-R_{6}}^{DB}+\overbrace{R_{14}-R_{13}}^{CBT}+\overbrace{R_{12}-R_{11}}^{VBT}\right]=0$$
(17-b)

$$\frac{d[DB^{+}]}{dt}=[R_{1}+R_{7}-R_{3}-R_{5}]=0 \qquad (17\text{-}c)$$

$$\frac{d[DB^{-}]}{dt}=[R_{2}+R_{8}-R_{4}-R_{6}]=0 \qquad (17\text{-}d)$$

By solving Eq. (17-c) and Eq. (17-d), one can obtain the occupation probabilities under illumination as:

$$\left.\begin{array}{l}F_{DB}^{+}=\dfrac{1}{1+\dfrac{\varepsilon_{p}^{+}+nS_{n}^{+}}{\varepsilon_{n}^{0}+pS_{p}^{0}}\left[1+\dfrac{\varepsilon_{p}^{0}+nS_{n}^{0}}{\varepsilon_{n}^{-}+pS_{p}^{-}}\right]}\\[2ex]F_{DB}^{0}=\dfrac{1}{1+\dfrac{\varepsilon_{n}^{0}+pS_{p}^{0}}{\varepsilon_{p}^{+}+nS_{n}^{+}}+\dfrac{\varepsilon_{p}^{0}+nS_{n}^{0}}{\varepsilon_{n}^{-}+pS_{p}^{-}}}\\[2ex]F_{DB}^{-}=1-F_{DB}^{0}-F_{DB}^{+}\end{array}\right\} \quad (18)$$

On the assumption that the localized states in the band tails are monovalent and obey Shockley-Read statistics in non-equilibrium condition, it can be shown that, for an arbitrary distribution $N_x(E)$, the steady state electron occupation rate $F_x(E)$ of a recombination centre located at energy $E$ is the same as originally derived by Shockley and Read for a single level.[87] Here, the subscript "x" can be replaced by CT or VT if the concerned states are CBT or VBT. Therefore, $F_{CT}(E)$ and $F_{VT}(E)$ can be represented as :

$$F_{CT}(E)=\frac{S_{n}^{CT}n+S_{p}^{CT}p^{'}}{S_{n}^{CT}(n+n^{'})+S_{p}^{CT}(p+p^{'})};$$

$$F_{VT}(E)=\frac{S_{n}^{VT}n^{'}+S_{p}^{VT}p}{S_{n}^{VT}(n+n^{'})+S_{p}^{VT}(p+p^{'})} \qquad (19)$$

### IV.B.4. Carrier densities in band tail states and in DB states

The carrier densities mentioned in Eq. (8) are given by the integrals over the states from the VB to the CB:

$$Q_{CT}=\int_{E_{V}}^{E_{C}}N_{CT}(E)F_{CT}(E)dE=Q_{CT}(n,p) \qquad (20)$$

$$Q_{VT}=\int_{E_{V}}^{E_{C}}N_{VT}(E)[1-F_{VT}(E)]dE=Q_{VT}(n,p) \qquad (21)$$

where $N_{CT}(E)$ and $N_{VT}(E)$ are the densities of states per unit energy of the CBT and VBT respectively.

$$N_{CT}(E)=N_{CT}^{0}\exp\left[-\frac{(E_{c}-E)}{kT_{c}}\right];$$

$$N_{VT}(E)=N_{VT}^{0}\exp\left[-\frac{(E-E_{v})}{kT_{v}}\right] \qquad (22)$$

$T_c$ and $T_v$ are the characteristic temperatures describing the slope of the exponentials of CBT and VBT respectively. Here CBT and VBT are assumed to be an exponential energy distribution.

The DB density of states located at energy $E_{DB}$ with total density $N_D$ is modeled by a Gaussian of width $W$:

$$N_{DB}(E)=\frac{N_{D}}{(2\pi)^{1/2}W}\exp\left[\frac{(E-E_{DB})^{2}}{2W^{2}}\right] \qquad (23)$$

While the total charge densities $Q_{DB}$ at DBs can be expressed as:

$$Q_{DB}=\int_{E_{V}}^{E_{C}}N_{DB}(E)(F_{DB}^{-}(E)-F_{DB}^{+}(E))dE=Q_{DB}(n,p) \qquad (24)$$

The excess photocarriers generated in DB states can be written as:

$$Q_{DB}(n,p)-Q_{DB}(n_{0},p_{0})=N_{DB}(F_{DB}^{0}+2F_{DB}^{-}-F_{DB}^{00}-2F_{DB}^{0-})$$

The expressions in Eq. (20), (21) and (23), are valid at thermal and under illumination. If the equilibrium density $n_0$ of free electrons and the equilibrium density $p_0$ of free holes are known, charge conservation between the equilibrium state and the illuminated state relates the density of one type of carrier to the other directly. Therefore, the charge neutrality condition between the equilibrium state and the illumination state gives

$$[n-n_{0}]-[p-p_{0}]+[Q_{CT}(n,p)-Q_{CT}(n_{0},p_{0})]-[Q_{VT}(n,p)-Q_{VT}(n_{0},p_{0})]+$$
$$N_{DB}(F_{DB}^{0}+2F_{DB}^{-}-F_{DB}^{00}-2F_{DB}^{0-})=0$$
(25)

### IV.B.5. The recombination equation

Considering the case of Eq. (17-a) where the rate of change, $dn/dt$, of electrons in the DB, CBT and VBT states are shown. The steady-state condition $dn/dt=0$ leads the continuity equation for electrons to be expressed in the following way:

$$G_{L}=\overbrace{R_{1}-R_{3}+R_{2}-R_{4}}^{DB\langle U_{DB}\rangle}+\overbrace{R_{9}-R_{10}}^{CBT\langle U_{CT}\rangle}+\overbrace{R_{16}-R_{15}}^{VBT\langle U_{VT}\rangle}$$

or $G_{L}=U_{CT}+U_{VT}+U_{DB}$ (26)

The net rates of recombination $U_{CT}$ and $U_{VT}$ on the CBT and VBT, respectively, are deduced from the basic equations of capture and emission processes of free electron [utilizing Eqs. (12) and (14)].

$$R_{9}=nS_{n}^{CT}N_{CT}(E)[1-F_{CT}(E)];$$
$$R_{10}=\varepsilon_{n}N_{CT}(E)F_{CT}(E)=n'N_{CT}(E)S_{n}^{CT}F_{CT}(E) \qquad (27\text{-}a)$$

$$R_{16}=nS_{n}^{VT}N_{VT}(E)[1-F_{VT}(E)];$$
$$R_{15}=\varepsilon_{n}N_{VT}(E)F_{VT}(E)=n'N_{VT}(E)S_{n}^{VT}F_{VT}(E) \qquad (27\text{-}b)$$

$$U_{CT} = R_9 - R_{10} = \int_{E_v}^{E_c} N_{CT}(E) \left[ \frac{S_n^{CT} S_p^{CT} (np - n_0 p_0)}{S_n^{CT}(n+n') + S_p^{CT}(p+p')} \right] dE \quad (28)$$

$$U_{VT} = R_{16} - R_{15} = \int_{E_v}^{E_c} N_{VT}(E) \left[ \frac{S_n^{VT} S_p^{VT} (np - n_0 p_0)}{S_n^{VT}(n+n') + S_p^{VT}(p+p')} \right] dE \quad (29)$$

Similarly, the recombination rates due to the DB centers can be deduced from the capture and emission processes of free electrons in those centers as shown in Eq. (11).

$$U_{DB} = \int_{E_v}^{E_c} N_{DB}(E) \left[ n(F_{DB}^+ S_n^+ + F_{DB}^0 S_n^0) - (F_{DB}^0 \varepsilon_n^0 + F_{DB}^- \varepsilon_n^-) \right] dE \quad (30)$$

Analogous expressions hold for the recombination rates if the case of free holes is considered:

$$U_{DB} = \int_{E_v}^{E_c} N_{DB}(E) \left[ p(F_{DB}^- S_p^- + F_{DB}^0 S_p^0) - (F_{DB}^0 \varepsilon_p^0 + F_{DB}^+ \varepsilon_p^+) \right] dE \quad (31)$$

However, throughout our calculations we will take only the case of free electrons. Hence, the net balance between the recombination rates (considering the Eqs. 28 - 30) and generation rate ($G_L$) can be expressed as:

$$G_L = \int_{E_v}^{E_c} N_{CT}(E) \left[ \frac{S_n^{CT} S_p^{CT}(np - n_0 p_0)}{S_n^{CT}(n+n') + S_p^{CT}(p+p')} \right] dE +$$
$$\int_{E_v}^{E_c} N_{VT}(E) \left[ \frac{S_n^{VT} S_p^{VT}(np - n_0 p_0)}{S_n^{VT}(n+n') + S_p^{VT}(p+p')} \right] dE +$$
$$\int_{E_v}^{E_c} N_{DB}(E) \left[ n(F_{DB}^+ S_n^+ + F_{DB}^0 S_n^0) - (F_{DB}^0 \varepsilon_n^0 + F_{DB}^- \varepsilon_n^-) \right] dE \quad (32)$$

### IV.B.6. Computing procedures

In modeling coplanar samples, it is a common practice to assume uniform electric field, perfect Ohmic contacts, and to neglect any transport driven by free-carrier diffusion. Under these assumptions, the continuity equation for holes is automatically satisfied when the continuity equation for electrons and the charge neutrality condition are fulfilled.

The kinetic steady state [Eq. (25)] and charge neutrality equations [Eq. (32)] are solved simultaneously by applying Newton-Raphson method for finding the roots of $n$ and $p$. We applied Simpson's method for numerical integration to find the integrals in Eqs. (20), (21), (24), (25) and (32). Once $n$ and $p$ were found for particular temperature and illumination intensity, we can obtain the value of photoconductivity by putting their values in Eq. (1) using physically justifiable values of mobilities. The code was written in Mathematica™ to be run on a desktop PC.

### IV.C. Selection of DOS and parameter values in μc-Si:H system

To construct a parameter space required for our computer simulations, our approach has been to employ the experimentally determined parameters wherever these are available. Further, reasonable and physically justifiable values of parameters were taken from the literature. The simulations were carried out to determine the photoconductivity as a function of temperature and illumination intensity, $\sigma_{ph}(T, G_L)$. As we are not using an optimization or fitting algorithm, the numerically determined $\sigma_{ph}(T, G_L)$ was compared with the experimental data graphically to reproduce the shape and magnitude of $\sigma_{ph}(T)$ and $\gamma(T)$. Initially we performed an extensive modeling study to understand the sensitivity of $\sigma_{ph}(T, G_L)$ on various parameters. The choice of various parameters as shown in Table II is briefly discussed below:

- A dangling bonds state distribution was incorporated in DOS to account for the presence of disorder phase associated with the boundary regions, similar to Tran's model [case-B1 of Ref. 50].
- Initially we tested our model by setting the mobility gap ($E_g = E_c - E_v$) of μc-Si:H as 1.8 eV, a value which is frequently used for a-Si:H. The logic behind using this value is that though the crystalline portion of the material affects the density of localized states in the gap, but it is the defect states located in the *disorder phase*, that actually take part in the recombination processes.[22] However, we also carried out calculations with a mobility gap of 1.12 eV for our fully crystallized μc-Si:H. We considered the DB distribution and its parameter values as mentioned by Lips *et al.* in their work regarding electron spin resonance studies on μc-Si:H.[20]
- Experimentally determined values for Fermi level positions were used (see Table I).
- The values of the carrier mobilities used in the simulation are those that we obtained from TRMC experiment, mentioned in Table I.
- For the values of the capture coefficients of the states (band tail states and DB states), we used the data published in various relevant works. It is known that the numerical values of the capture coefficients do affect phototransport properties,[43,95] but our aim here is only to find a scientifically acceptable value for our simulation, and not actually attempt to standardize the values for the μc-Si:H system. An analysis of the literature on the role of the value of capture coefficients in the phototransport properties helped us to decide which values would be applicable in our case. The work by Tran[50]

Table II. Details of parameter values used in the model simulations.

| Types, Model Cases & $E_g$ | $\mu_n$ | $\mu_p$ | $N_D$ | Valence Band Tails (VBT) ||||||||| Conduction Band Tails (CBT) ||||||
|---|---|---|---|---|---|---|---|---|---|---|---|---|---|---|---|---|
| | | | | Shallower VBT (VBT$_1$) Eq. (33) ||| Deeper VBT (VBT$_2$) Eq. (34) |||| Shallower CBT (CBT$_1$) Eq. (35) ||| Deeper CBT (CBT$_2$) Eq. (36) |||||
| | | | | $E_{tv1}$ | $kT_{v1}$ | $kT_{v2}$ | $E_{v1}$-$E_v$ | $N_{v2}^0$ | $E_{tv2}$ | $kT_{v3}$ | $kT_{v4}$ | $E_{tc1}$ | $kT_{c1}$ | $kT_{c2}$ | $E_{c1}$-$E_v$ | $N_{ct2}^0$ ($10^{14}$) | $E_{tc2}$ | $kT_{c3}$ | $kT_{c4}$ |
| Type-A | | | | | | | | | | | | | | | | | | | |
| A1(1.8) | 1 | 0.1 | $10^{15}$ | 0.2 | 0.1 | 0.05 | 0.44 | $10^{17}$ | 0.4 | 0.1 | 0.05 | 0.18 | 0.08 | 0.024 | 1.375 | 1.8 | 0.35 | 0.13 | 0.03 |
| A2(1.8) | 1 | 0.1 | $10^{15}$ | 0.2 | 0.1 | 0.05 | 0.593 | $10^{15}$ | 0.4 | 0.2 | 0.05 | 0.18 | 0.08 | 0.024 | 1.375 | 1.8 | 0.35 | 0.13 | 0.03 |
| Type-B | | | | | | | | | | | | | | | | | | | |
| B1(1.8) | 5 | $10^{-3}$ | $10^{14}$ | 0.125 | 0.04 | 0.012 | 0.223 | $10^{15}$ | 0.25 | 0.07 | 0.025 | 0.15 | 0.1 | 0.02 | 1.522 | 1000 | 0.15 | 0.06 | 0.03 |
| B2(1.8) | 5 | 0.1 | $10^{14}$ | 0.125 | 0.04 | 0.012 | 0.223 | $10^{15}$ | 0.25 | 0.07 | 0.025 | 0.125 | 0.05 | 0.02 | 1.595 | 3000 | 0.12 | 0.12 | 0.03 |
| B3(1.12) | 5 | 0.1 | $10^{14}$ | 0.125 | 0.04 | 0.012 | 0.223 | $10^{15}$ | 0.25 | 0.07 | 0.025 | 0.125 | 0.05 | 0.02 | 0.915 | 3000 | 0.12 | 0.12 | 0.03 |
| Type-C | | | | | | | | | | | | | | | | | | | |
| C1(1.8) | 10 | 0.5 | $10^{14}$ | 0.125 | 0.04 | 0.012 | 0.181 | $10^{17}$ | 0.35 | 0.07 | 0.025 | 0.125 | 0.07 | 0.005 | -- | -- | -- | -- | -- |
| C2(1.8) | 10 | 1 | $10^{14}$ | 0.125 | 0.04 | 0.012 | 0.181 | $10^{17}$ | 0.35 | 0.07 | 0.025 | 0.125 | 0.03 | 0.005 | 1.624 | 1 | 0.3 | 0.25 | 0.04 |
| C3(1.12) | 10 | 1 | $10^{14}$ | 0.125 | 0.04 | 0.012 | 0.181 | $10^{17}$ | 0.35 | 0.07 | 0.025 | 0.125 | 0.03 | 0.005 | 0.944 | 1 | 0.3 | 0.25 | 0.04 |

Units: $N_{v2}^0$, $N_{ct2}^0$ -(cm$^{-3}$eV$^{-1}$); $E_g$ (=$E_c$- $E_v$), ($E_{v1}$ - $E_v$), ($E_{c1}$- $E_v$) - (eV); $E_{tv1}$, $E_{tv2}$, $E_{tc1}$, $E_{tc2}$ - (eV); $kT_{v1}$, $kT_{v2}$, $kT_{v3}$, $kT_{v4}$, $kT_{c1}$, $kT_{c2}$, $kT_{c3}$, $kT_{c4}$ - (eV); $\mu_n$, $\mu_p$ – (cm$^2$/V-s); $N_D$ -(cm$^{-3}$)

Fixed Parameter values: $N_{v1}^0 = N_{ct1}^0 = 10^{21}$ cm$^{-3}$eV$^{-1}$; $E_v = 0$ eV;

*For dangling bonds and capture cross section values see §IV.C; For Fermi level position of these set of samples see Table I*

has comprehensive description regarding capture coefficients in *a*-Si:H, which is pertinent to our study because *μ*c-Si:H and *a*-Si:H have some common overlapping features, especially when we consider the DBs states in the gap of our material. It is pragmatic and acceptable to use the knowledge of DBs in *a*-Si:H, because there is no model available as yet for DBs in *μ*c-Si:H. Balberg *et al.*[22] have also demonstrated the role of choice of capture coefficient values for *single phase μc*-Si:H. In addition, we also considered few recent experimental studies mentioning the values of capture coefficient in *μ*c-Si:H before we decided on a set of standard values, which we used in our simulation model.[24,25,52]

A set of initial parameters used as a standard for all the three cases is given below. For sensitivity analyses, some of the parameters were altered to improve the quantitative agreement of the simulation results with the experimental data. Though we proceeded further in each case by analyzing the flaws and errors in every model that did not yield the desired results, but for the sake of brevity, we have presented here only the suitable results that we deemed acceptable and physically correct.

▪ **Dangling Bond States:**

*For $E_c$- $E_v$ = 1.8 eV,*

The DBs distribution, its energetic position and the values of its parameters were considered to be the same as Tran's model (Case-B1, in Ref. 50):

$$N_{DB}(E) = \frac{N_D}{(2\pi)^{1/2}W} \exp\left[\frac{(E-E_{DB})^2}{2W^2}\right]; \text{DB is located at energy } E_{DB} \text{ and } W \text{ is the Gaussian width. } E_{DB} - E_v = 0.75$$

eV, $W$= 0.15 eV, and $U$ =0.4 eV.

*For $E_c$- $E_v$ = 1.12 eV,*

We considered the work of Lips *et al.*,[20] which has suggested two types of DBs, '*db1*' and '*db2*', having different distributions within the gap. In our simulation, we have considered their '*db2*', the dangling bond which plays a important role in the recombination process as it has a strong dependence on the Fermi level position, whereas their '*db1*' does not. The estimated Gaussian width, $W$, of the defect distribution located at energetic position, $E_{DB}$, with positive correlation energy $U$: $E_{DB} - E_v$ = 0.3 eV, $W$ = 0.3 eV, and $U$ = 0.3 eV.

For both the cases, we have taken a low value of the DB defect density; $N_D$, was taken to be $10^{14}$ cm$^{-3}$ as our material has a low concentration of DBs.

▪ **Band Tail States**

At the band edges the tail states have been variously claimed to be slightly deviated, or linear,[89,96] and even parabolic, in both *a*-Si:H and *μ*c-Si:H. We have also not considered the tail states at the edges to be exponential, instead, the slope of a tail state is considered to change with the changing energy level in the gap. It is not feasible to use such tail state distributions in calculations. Therefore, we used formulae and altered the variables to attain the desired shape of tail states distribution,[97] as determined by our qualitative and experimental analyses. The formulae used to obtain such tail state distributions in the conduction and valence band regions are given as following:

**The shallower valence band tail ($VBT_1$) is given by:**

$$N_{VT1} = N_{VT1}^0 \times \exp\left[-\frac{(E-E_v)}{kT_{v1}}\right] \times \left[\frac{\exp\left[\frac{(-(E-E_v)+E_{tv1})}{kT_{v2}}\right]}{1+\exp\left[\frac{(-(E-E_v)+E_{tv1})}{kT_{v2}}\right]}\right]$$

(33)

**The deeper valence band tail ($VBT_2$) is given by:**

$$N_{VT2} = N_{VT2}^0 \times \exp\left[-\frac{(E-E_{v1})}{kT_{v3}}\right] \times \left[\frac{\exp\left[\frac{(-(E-E_{v1})+E_{tv2})}{kT_{v4}}\right]}{1+\exp\left[\frac{(-(E-E_{v1})+E_{tv2})}{kT_{v4}}\right]}\right]$$

(34)

**The shallower conduction band tail ($CBT_1$) is given by:**

$$N_{CT1} = N_{CT1}^0 \times \exp\left[-\frac{(E_c-E)}{kT_{c1}}\right] \times \left[\frac{\exp\left[\frac{-((E_c-E)-E_{tc1})}{kT_{c2}}\right]}{1+\exp\left[\frac{-((E_c-E)-E_{tc1})}{kT_{c2}}\right]}\right]$$

(35)

**The deeper conduction band tail ($CBT_2$) is given by:**

$$N_{CT2} = N_{CT2}^0 \times \exp\left[-\frac{(E_{c1}-E)}{kT_{c3}}\right] \times \left[\frac{\exp\left[\frac{-((E_{c1}-E)-E_{tc2})}{kT_{c4}}\right]}{1+\exp\left[\frac{-((E_{c1}-E)-E_{tc2})}{kT_{c4}}\right]}\right]$$

(36)

$N_c = N_{CT1}^0 kT$ and $N_v = N_{VT1}^0 kT$ being the effective density of states at $E_c$ and $E_v$ respectively.

▪ **Capture coefficients:**

$S_n^{CT} = 10^{-9} (\text{cm}^3\text{s}^{-1})$; $S_p^{CT} = 10^{-15} (\text{cm}^3\text{s}^{-1})$;

$S_p^{VT} = 10^{-6} (\text{cm}^3\text{s}^{-1})$; $S_n^{VT} = 10^{-8} \left(\frac{T}{300}\right)^3 (\text{cm}^3\text{s}^{-1})$;

$S_n^+ = S_p^- = 3\times 10^{-8} (\text{cm}^3\text{s}^{-1})$; $S_n^0 = S_p^0 = 3\times 10^{-9} (\text{cm}^3\text{s}^{-1})$.

The values of capture coefficients for electrons and holes of shallower tail states ($CBT_1$ or $VBT_1$) were taken as mentioned above. However, the capture coefficients values for both the carriers in the deeper tail states ($CBT_2$ or $VBT_2$) were chosen one order less than their shallower tail states values.

It is worth mentioning here that though tunneling is considered important at low temperatures in *a*-Si:H[98,99] as well as in *μ*c-Si:H system,[77,100] we have not considered it in our simulation. Tunneling processes consist of tunneling conduction and tunneling recombination between localized states. First, considering the tunneling conduction processes, it is known that the constant behavior of $\sigma_{ph}$ at low *T* is explained by energy-loss tunneling among localized states.[101,102] We also found $\sigma_{ph}$ to be independent of temperature when *T* < 50 K in our SSPC experiments, in all the three types of material, which can only be explained by the occurrence of tunneling processes. It can be mentioned here that the occurrence of this effect is also corroborative with our low temperature dark conductivity experimental results. However, this process is important only at temperatures below 100 K, while our interest lies in the temperature range above 200 K, where the important phototransport processes, like thermal quenching and temperature dependent *γ* behavior, are observed. This temperature range under study, thus, is well above the temperature where the tunneling conduction becomes more dominant than the free carrier conduction, and in our case, for all practical purposes, tunneling conduction by electrons can be neglected.

The other important tunneling process is the tunneling recombination of nonequilibrium carriers captured by localized states, in which, it is believed that in a heterogeneous system like poly-Si or *μ*c-Si:H, tunneling occurring under the potential barriers that separate the carriers determines the recombination rate.[77,100] Besides, in *a*-Si:H or any system where two tail states and dangling bond states are considered, there is occurrence of tail-tail and/or tail-DB tunneling.[98] Radiative tail-tail recombination is dominant below 40 K, and decreases with increasing temperature above 80K. Again, this phenomenon will not have any role in the temperature range in which we are interested. Though it is generally agreed that incorporation of tunneling processes decreases the discrepancy between experimental data and the simulation results, but the mechanism of tunneling recombination does not change our qualitative understanding of thermal quenching or other phototransport properties.

### IV.D. Results of SSPC modeling study

In our study, we have meticulously applied many possibilities before considering the results that were consistent with experimental findings associated with a particular *type* of *μ*c-Si:H material. Since each type of material has a different phototransport behavior and a different DOS profile, necessitating separate simulation models, we have presented the results separately for each type of film microstructure. Our preliminary attempts at numerical modeling had yielded acceptable results that we had published earlier. But our continued efforts directed towards fine-tuning the DOS features further, have since then, generated many results that are more accurate, with the simulated values much closer to the experimental data. What we have presented here consists of some results that we had obtained earlier, and some that have been refined with extensive modeling.

### IV.D.1. Model simulation for Type-A *μc*-Si:H (0.5 < *γ* < 1 with a TQ effect)

We already have the information regarding at least one of the tail state distribution in the vicinity of CB edge for this material (§ III.D.1). The detailed study of electrical transport properties of *type-A μ*c-Si:H films has revealed some similarities to the properties of *a*-Si:H. Therefore, for the selection of DOS distribution in the VB region of this type of *μ*c-Si:H, we assigned such values of parameters that would yield an overall slope of the structured VBT profile similar to that seen in amorphous silicon. Device quality *a*-Si:H has a VBT with a width of 0.045 eV and a deep defect density (especially DBs) in the range of $10^{15} - 10^{16}$ cm$^{-3}$. We would like to mention here that we also tried simulation with an exponential VBT having a single slope of the same width, but this did not generate phototransport behavior corroborative with the experimental values. We have presented only the relevant results that we considered acceptable.

**Case-A1**

Considering the mobility gap, $E_c - E_v = 1.8$ eV, the constructed effective DOS distribution of *type-A μ*c-Si:H material is shown in Fig. 8(a). Here we see that we have used the CBT as derived from the experimental data, and a structured VBT, that has been split into a shallow and a deeper tail states. The dark conductivity results predict the Fermi level position to be around ≈ 0.46 eV below CB edge, i.e., $E_f = E_c - 0.46$ eV (Table I). The density of DBs has been taken ≈ $10^{15}$ cm$^{-3}$ (Table II).

Using this effective DOS profile, we obtained the simulated temperature dependent $\sigma_{ph}$ and light intensity exponent for this *type-A* material, which are shown in Fig. 8(b) and (c) respectively. It can be seen in Fig. 8(b) that the simulated photoconductivity increases monotonically with the increase in temperature. However, at temperature above 250 K, the monotonic increment of $\sigma_{ph}$ does not slow down with further rise in temperature [*TQ* behavior of $\sigma_{ph}$ (*T*)], as would be expected from the experimental observations [Fig. 1 (a) of §III.C], except for low light intensity conditions. Now, looking at the temperature dependent behavior of light intensity exponent, *γ* in Fig. 8 (c), we see that *γ* initially decreases monotonically with increasing temperature, followed by a rise at temperature above 250K, though not as marked as we found in our experimental results [Fig. 1(c) of §III.C]. We further used the same DOS profile with some changes in the parameters like DB density, Fermi level position, etc., but these failed to reproduce the experimental results. We have not presented the results of these trial cases.

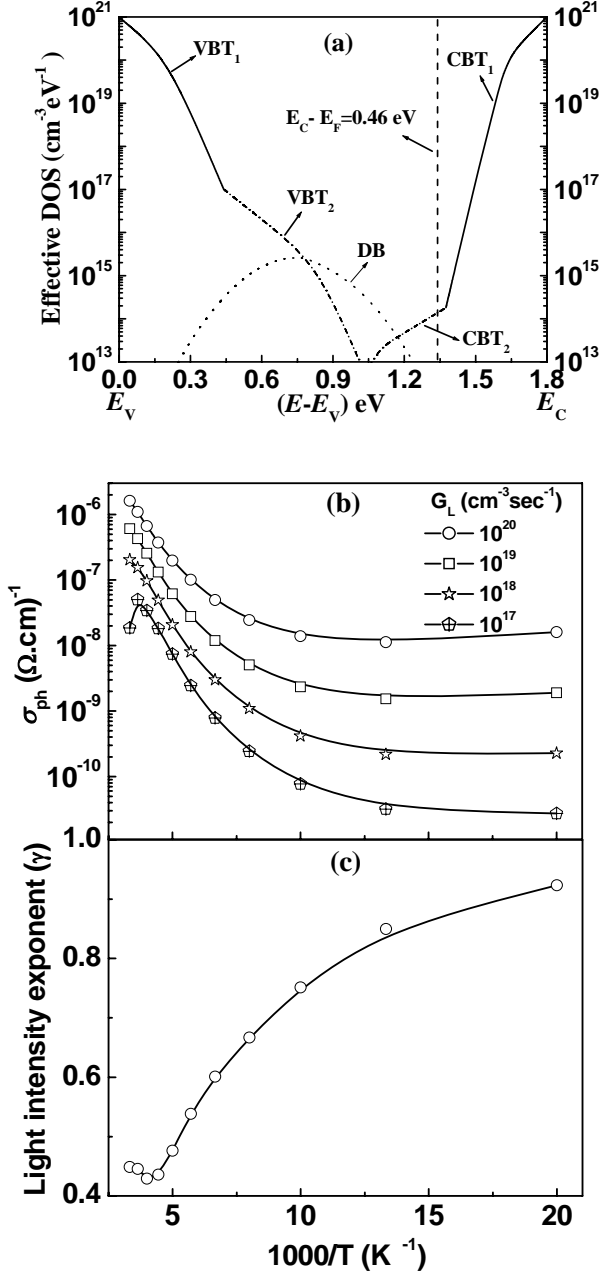

FIG. 8. (a) The effective DOS model for *type-A* μc-Si:H material corresponding to Case-A1; (b) simulated temperature dependent $\sigma_{ph}(T)$ for various light intensities and (c) simulated temperature dependent light intensity exponent $\gamma(T)$.

## Case-A2

Next, we considered a *equilibrium* DOS profile as shown in Fig. 9(a), with its deeper $VBT_2$ states assigned a value ($10^{15}$ cm$^{-3}$ eV$^{-1}$) less than what we used in the previous case ($10^{17}$ cm$^{-3}$ eV$^{-1}$). The results of this model simulation are presented in Fig. 9(b) and (c). The temperature dependent $\sigma_{ph}(T)$ in Fig. 9(b) exhibits the presence of *TQ* behavior at temperatures above 250 K, be-sides the monotonic increase with increase in temperature at lower *T* values, which is similar to the experimental results.

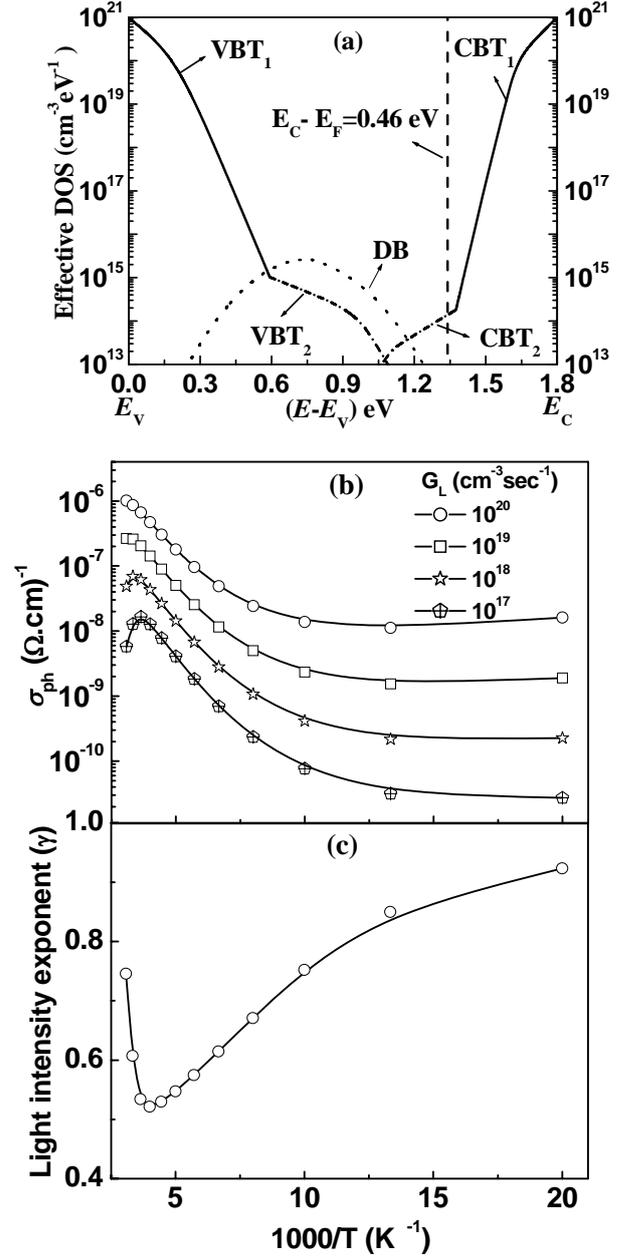

FIG. 9. (a) The proposed effective DOS model for *type-A* μc-Si:H material corresponding to Case-A2; (b) simulated temperature dependent $\sigma_{ph}(T)$ for various light intensities and (c) simulated temperature dependent light intensity exponent $\gamma(T)$.

The simulated temperature dependence of $\gamma$, as shown in Fig. 9 (c), exhibits monotonic decrease with the increase in temperature. However, at higher temperature (above 250 K), it rises sharply due to the thermalization of carriers, a behavior which is very close to what we observed in our experimental findings [Fig. 1(c) of §III.C].

Therefore, simulation using this type of VBT yields results of the phototransport properties of *type-A* μc-Si:H material that match well with the experimental data.

### IV.D.2. Model simulation for Type-B μc-Si:H (0.5 < γ < 1 with no TQ effect)

Similar to the procedure followed in *type-A* material, here also we assigned the experimentally determined values to the CBT in the simulation. For VBT, we assigned the numerical values that we had qualitatively derived in § III.D.4, taking into account a structured appearance, with a low value of density of deeper tail states. In contrast to the *type-A* material, the shallower VBT in this type of material has a steeper slope.

**Case-B1**

The effective DOS distribution of *type-B* material is presented in Fig. 10(a). Here we have considered the band gap, $E_c - E_v = 1.8$ eV. The dark conductivity results predict the Fermi level position to be around ≈ 0.42 eV below CB edge, i.e. $E_f = E_c - 0.42$ eV (Table I). The details of parameter values are given in Table II.[24]

The simulated temperature dependent photoconductivity of this type of material is shown in Fig. 10(b). The simulated $\sigma_{ph}(T)$ increases monotonically with the increase in temperature without any *TQ* effect. Similar observation was also found in the experimental results of temperature dependent photoconductivity of this material [Fig. 2(a)]. However, at lower temperatures, below 125 K, a rise in $\sigma_{ph}$ can be seen, whereas the experimental results show the monotonic decrease sustained up to temperatures as low as 50 K. Here, we would like to mention that for a better quantitative agreement with the experimental $\sigma_{ph}$ values, it was essential to use $\mu_n = 5$ cm$^2$/V-sec and $\mu_p = 0.001$ cm$^2$/V-sec. This high ratio of $\mu_n / \mu_p$ has been reported in literature, and it implies that it is the mobility value that makes μc-Si:H an n-type photoconductor.

The simulated temperature dependence of γ is shown in Fig. 10 (c). γ is also seen to increase with decreasing temperature, a behavior similar to what we observed in our experimental findings [Fig. 2(c)]. Nevertheless, in the experimental results, the initial rise in γ is steep up to a temperature ≈100 K, where it gradually starts slowing down, showing saturation. In contrast, the simulated γ shows an exaggerated kink at ≈125 K, instead of a gradual saturation effect.

**Case-B2**

Therefore, considering the simulation results of both the phototransport properties obtained from Case-B1, we found the DOS profile of this material somewhat less than satisfactory, and proceeded to fine-tune it further by altering the numerical values of the CBT, keeping the slope at every energetic position the same as we derived from our experiments. In this trial simulation, we did not make any other alterations, so that we could isolate the effect of the change in the CBT features. The DOS profile thus obtained is shown in Fig. 11(a). The phototransport results using this model simulation are presented in Fig. 11(b) and (c). It is evident that now the simulated results agree well with their experimental counterparts, the

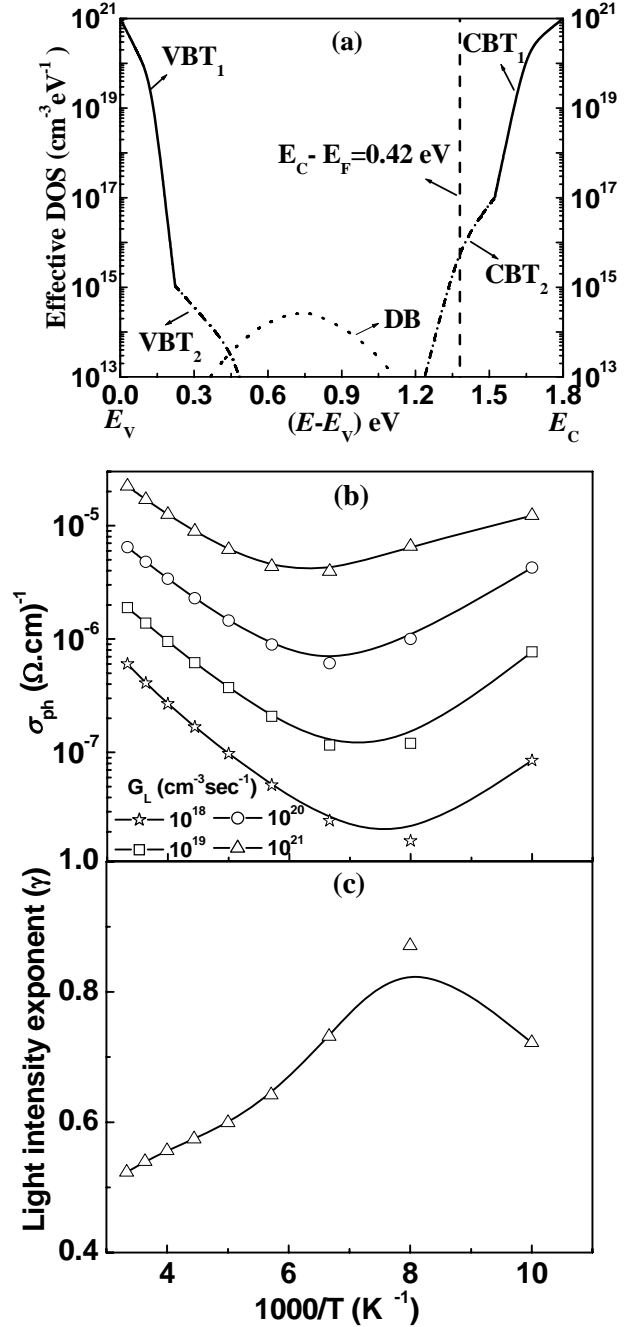

FIG. 10. (a) The effective DOS model for *type-B* μc-Si:H material corresponding to Case-B1 (band gap, $E_c-E_v =1.8$ eV); (b) simulated temperature dependent $\sigma_{ph}(T)$ for various light intensities and (c) simulated temperature dependent light intensity exponent γ(T).

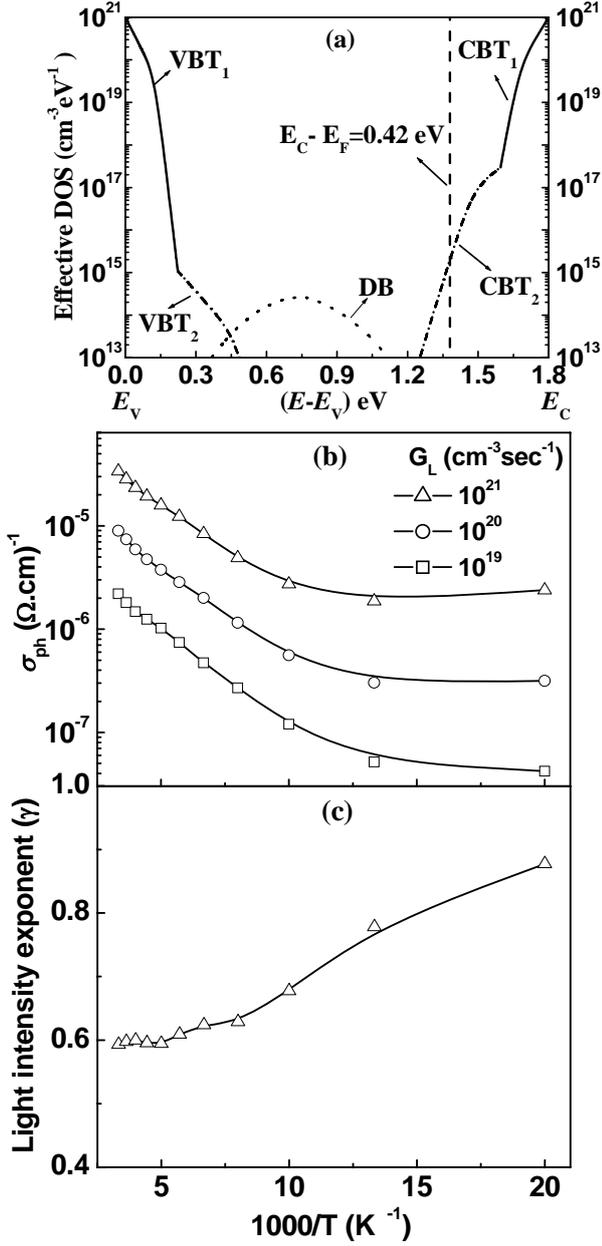
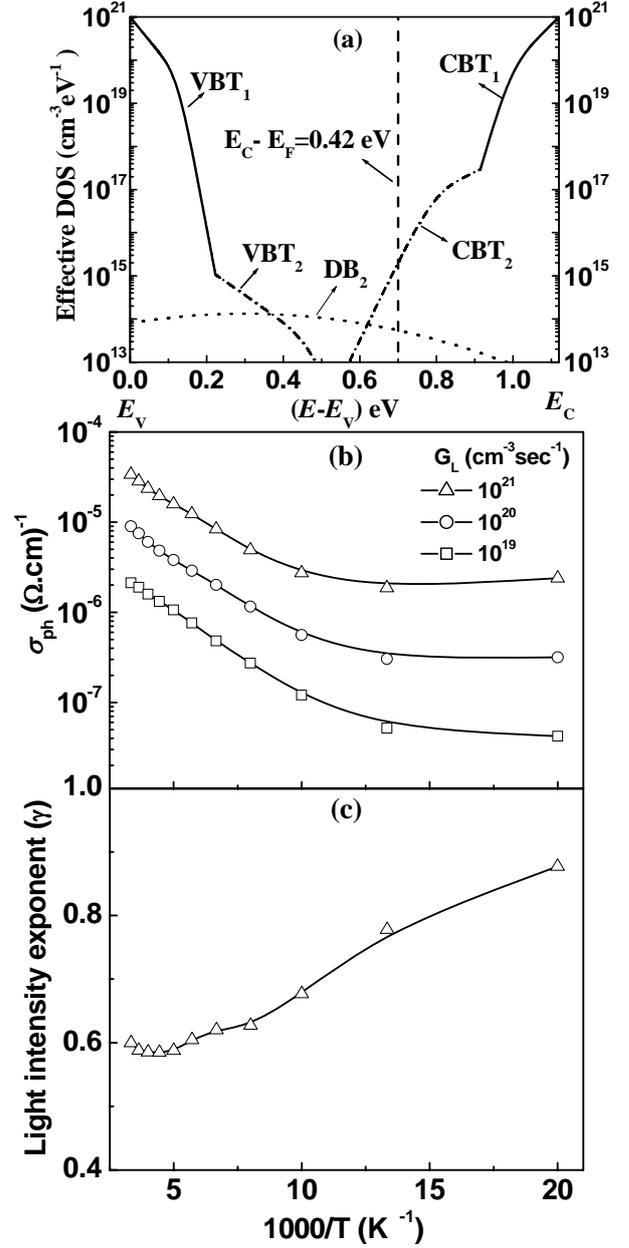

FIG. 11. (a) The proposed effective DOS model for *type-B* $\mu$c-Si:H material corresponding to Case-B2 (band gap, $E_c$-$E_v$ =1.8 eV); (b) simulated temperature dependent $\sigma_{ph}$ ($T$) for various light intensities and (c) simulated temperature dependent light intensity exponent $\gamma$ ($T$).

FIG. 12. (a) The effective DOS model for *type-B* $\mu$c-Si:H material considering the band gap, $E_c$-$E_v$ =1.12 eV (Case-B3); (b) simulated temperature dependent $\sigma_{ph}$ ($T$) for various light intensities and (c) simulated temperature dependent light intensity exponent $\gamma$ ($T$).

$\sigma_{ph}$ ($T$) showing a monotonic decrease with decreasing temperature well down to 50 K [Fig. 11(b)], and $\gamma$ showing a gradual rise with decreasing temperature, without any kink [Fig. 11(c)], which was observed in the previous case [Fig. 10(c)]. In this case, we have chosen $\mu_n$ = 5 cm$^2$/V-sec and $\mu_p$ = 0.1 cm$^2$/V-sec, and thus the $\mu_n$ / $\mu_p$ ratio is more reasonable and closer to the reported values. We considered this effective DOS profile to be correct, acceptable and conforming excellently to our analytical approach.

### Case-B3

The structural properties of *type-B* material have been shown to have very good crystallinity in our structural investigations. One would expect the mobility gap in such a material to be somewhat on the lower side, rather than being similar to the higher value found in $a$-Si:H. Various modeling studies involving poly-Si materials, and $\mu$c-Si:H materials showing properties similar to poly-Si, have considered the mobility gap to be 1.12 eV, which is the known value for the gap of c-Si. Considering this value to be one end of the spectrum of mobility gap acceptable for a case of $\mu$c-Si:H material, we were interested to see how the phototransport properties would change if a mobility gap is reduced from 1.8 to 1.12eV for model simulation of this *type-B* $\mu$c-Si:H material.

The proposed effective DOS using the band gap of 1.12 eV is shown in Fig. 12(a). The results of phototransport properties obtained from this model DOS feature are illustrated in Fig. 12(b) and (c). Interestingly, the results of this model simulation are almost the same as what we found in the previous case using a gap of 1.8 eV. This will be discussed in detail in the coming section.

### IV.D.3. Model simulation for Type-C $\mu$c-Si:H (0.15 < $\gamma$ < 1 with a TQ effect)

In § III.D.3, we have qualitatively asserted that the effective DOS in fully crystallized $\mu$c-Si:H films of *type-C* may exhibit two different valence band tails; a sharper, shallow tail originating from grain boundary defects and another less steeper, deep tail associated with the defects in the columnar boundary regions. This theory of two VBT configurations explains best the phenomenon of sublinearity without precluding any of the other conditions for thermal quenching.

**Case-C1**

The proposed effective DOS distribution for this material is shown in Fig. 13(a) for a band gap $(E_c - E_v) = 1.8$ eV. The values of the parameters and formulae used to obtain such tail state distributions in the gap regions are mentioned in Table II. The dark conductivity results predict the Fermi level position to be around $\approx 0.34$ eV below CB edge, i.e., $E_f = E_c - 0.34$ eV (Table I).[24,25] Figure 13(b) shows the simulated $\sigma_{ph}$ as a function of temperature for different intensities of light. There is a monotonic increase in the $\sigma_{ph}$ with an increase in temperature. However, at higher temperatures above 250 K, the $\sigma_{ph}$ starts decreasing with further rise in temperature, thus exhibiting thermal quenching (*TQ*) in our simulated $\sigma_{ph}$ (*T*) as well. The onset of *TQ*, and its shift towards higher temperatures with increase in illumination, is the same as seen in our experimental results [Fig. 3(a)].

Now let us turn to the simulated temperature dependence of $\gamma$ for *type-C* material, which is shown in Fig. 13(c). The $\gamma$ value has a dip in its temperature dependence, reaching below 0.5 in the temperature region where *TQ* is observed, which closely resembles our experimental findings [Fig. 3(c)]. The simulation results of phototransport properties are seen to be in excellent agreement with their experimental findings, by considering $\mu_n = 10$ cm$^2$/V-sec and $\mu_p = 0.5$ cm$^2$/V-sec, which are close to the mobility values obtained for polycrystalline material. In spite of the overall satisfactory results, $\sigma_{ph}$ below 100 K as seen in the inset of Fig. 13(b) is somewhat deviating from the experimental results.

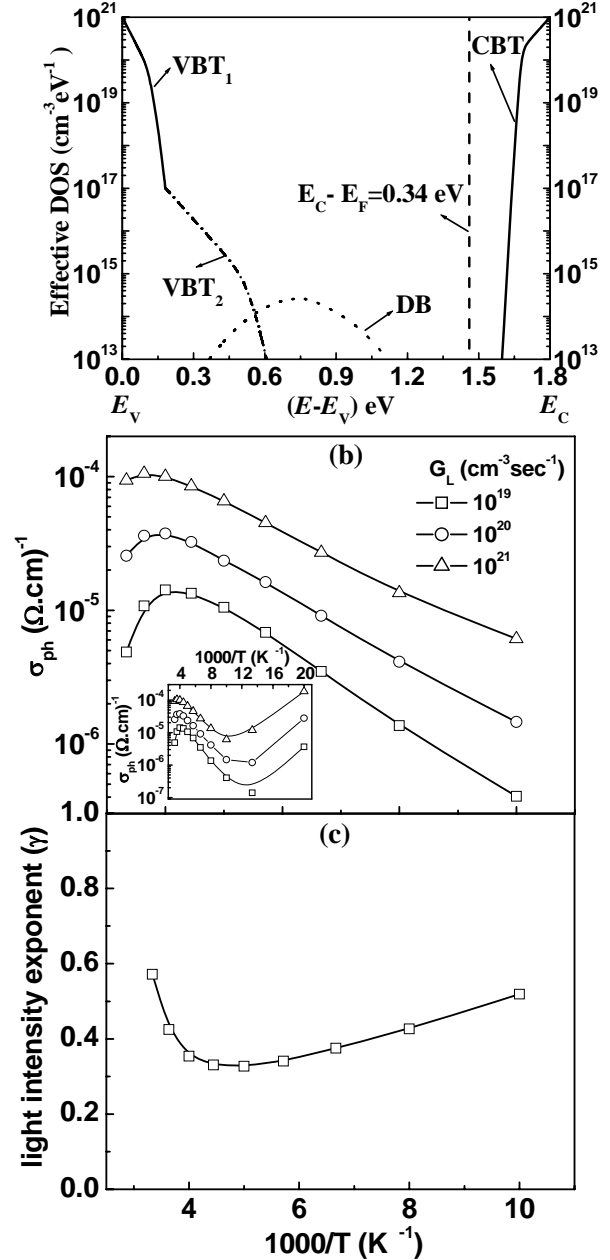

FIG. 13. (a) The effective DOS model for *type-C* $\mu$c-Si:H material considering the band gap, $E_c - E_v = 1.8$ eV (Case-C1); (b) simulated temperature dependent $\sigma_{ph}$ (*T*) for various light intensities and (c) simulated temperature dependent light intensity exponent $\gamma$(*T*).

The results of model simulation of *type-C* $\mu$c-Si:H using the DOS shown in Fig. 13(a), i.e., Case-C1, that are close to our experimental observation, motivated us to instigate a sensitivity analysis so that we can generate such simulated phototransport properties, which would be in conformity with the experimental results for the full range of temperature (experimental range). We tried to vary the position of $E_f$ and the quantitative values of the parameters of VBT and CBT. Interestingly, none of the changes could give us results better than what we saw in Fig. 13(b) and (c), except for an introduction of a deeper CBT$_2$ with a peak density of $10^{14}$ cm$^{-3}$eV$^{-1}$ with a slowly decaying tail states in the deeper side of the gap region.

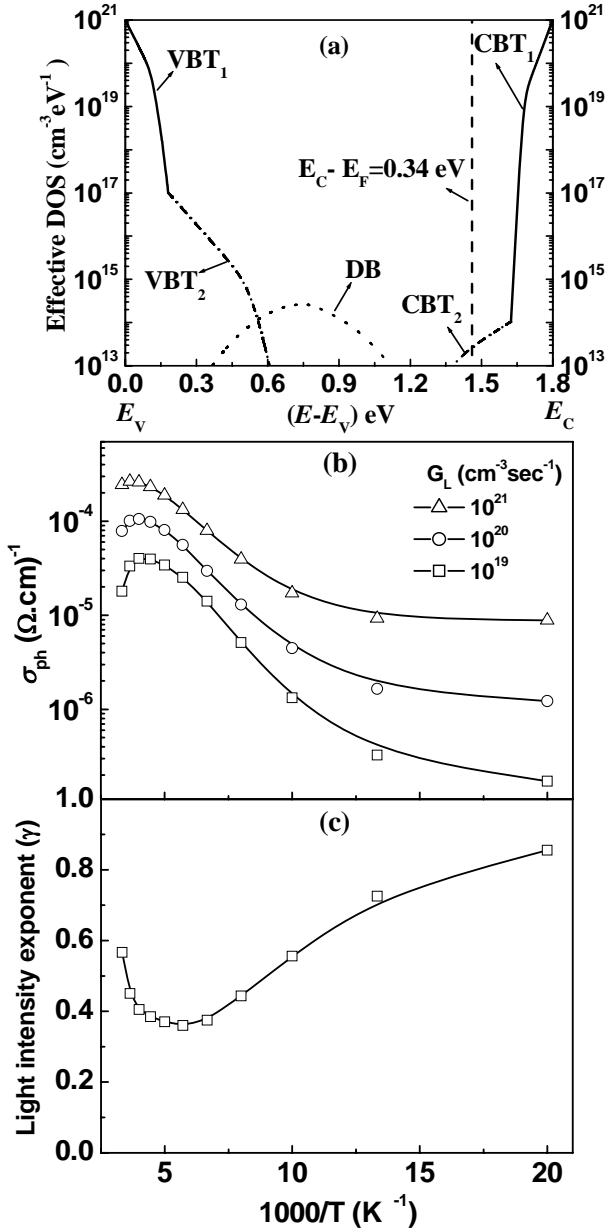

FIG. 14. (a) The proposed effective DOS model for *type-C* $\mu$c-Si:H material by considering the band gap, $E_c$-$E_v$ =1.8 eV (Case-C2); (b) simulated temperature dependent $\sigma_{ph}$ (T) for various light intensities and (c) simulated temperature dependent light intensity exponent $\gamma$(T).

**Case-C2**

The figure of this modified DOS profile is presented in Fig. 14(a). The behavior of simulated $\sigma_{ph}$ (T) obtained in this model simulation illustrated in Fig. 14(b) shows an excellent resemblance to the experimental findings. Unlike the $\sigma_{ph}$ (T) of previous case, the $\sigma_{ph}$ (T) thus obtained is consistent with the experimental results even down to 50 K. The light intensity exponent as shown in Fig. 14(c) on the other hand, shows somewhat better agreement than the one obtained in the previous case. In this case we considered $\mu_n$ = 10 cm$^2$/V-sec and $\mu_p$ = 1 cm$^2$/V-sec.

**Case-C3**

The structural investigations have revealed this material to be fully crystallized with tightly packed large crystallites without any trace of amorphous content. The electrical properties of *type-C* $\mu$c-Si:H are closer to those of polycrystalline type material. Just as we had tried to see the effect of band gap reduction in model simulation of *type-B* $\mu$c-Si:H, we tried the same here as well. The effective DOS model for the gap $E_c$-$E_f$ = 1.12 eV is shown in Fig. 15(a). The phototransport properties obtained by using this model are shown in Fig. 15(b) and (c) respectively for $\sigma_{ph}$(T) and $\gamma$(T). Thus, the figures 15(b) and (c) have a striking similarity, both qualitatively and quantitatively that will be discussed further in the next section.

### IV.E. Analysis of SSPC model simulation results

The results of simulated phototransport properties of all the three types of $\mu$c-Si:H having different microstructural properties show an excellent quantitative and qualitative agreement with their experimental observations. However, the motivation behind this modeling study was not to try to reproduce the experimental results in the simulation, but to understand the phototransport processes occurring in our constructed DOS distribution, which would be capable of producing phototransport behavior similar to our experimental observations, for any particular type of $\mu$c-Si:H material.

An interesting outcome of this study was the striking similarity observed between the results of model simulations based on the DOS distributions constructed for two different energetic gap values (1.12 eV and 1.8 eV). The band gap of the $\mu$c-Si:H system has not been studied previously in detail, though values between that of c-Si (i.e., 1.1 eV) and that of *a*-Si:H (i.e., 1.8 eV) have been reported. The results of *type-B* and *type-C* materials substantiate our previous argument (§IV.C) that it is the distribution of localized states in the gap, rather than the gap width itself, that affects the recombination processes, and hence the phototransport properties. The crystalline portion of the material is supposed to affect the distribution of localized tail states within the energetic gap, and their density in the gap. However, it is the disordered phase with the defect states located within, that actually influences the recombination processes, no matter where the disorder phase exists within the material (e.g., columnar or grain boundaries). Thus, keeping these arguments in view, we shall discuss the recombination processes obtained only from the model where $E_g$ = 1.8 eV, further ahead in this section.

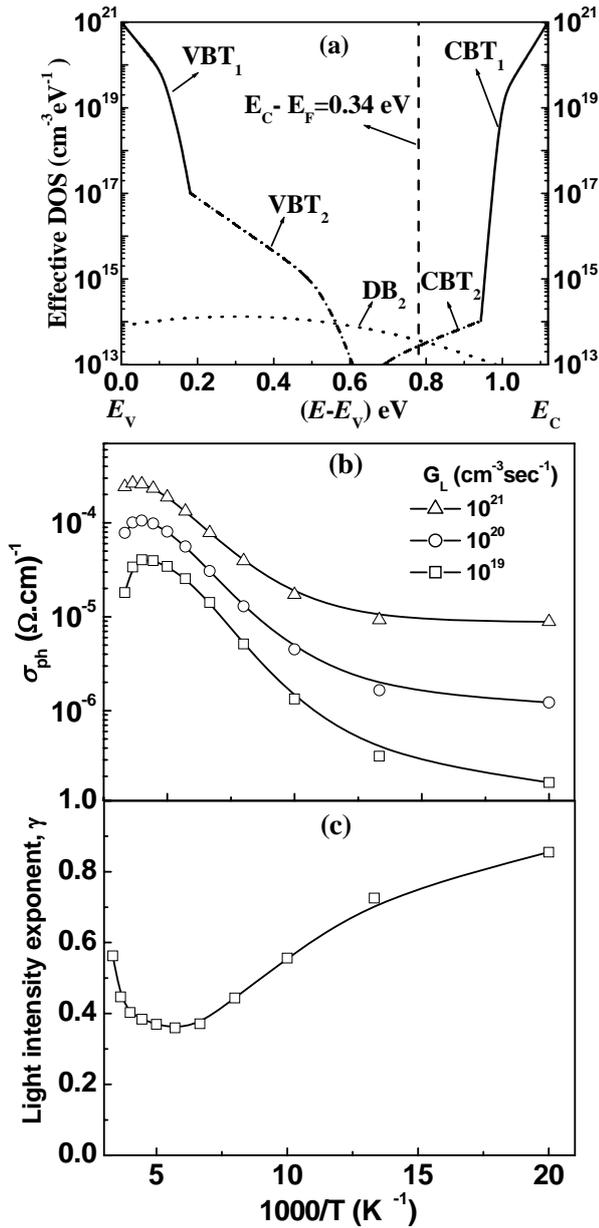

FIG. 15. (a) The effective DOS model for *type-C* μc-Si:H material considering the band gap, $E_c$-$E_v$ =1.12 eV (Case-C3); (b) simulated temperature dependent $\sigma_{ph}$ (T) for various light intensities and (c) simulated temperature dependent light intensity exponent γ (T).

Another interesting observation was regarding the role of DBs. In the case with $E_g$ = 1.8 eV, the DB distribution was considered to be narrow (Gaussian width W =0.15 eV), as was used by Tran.[50] A broader DB distribution (W =0.3 eV) was considered in the case with $E_g$ = 1.12 eV, as has been mentioned by Lips et al. in their work regarding electron spin resonance studies on μc-Si:H.[20] It appears from the similar results of the two cases that a change in the DB distribution hardly makes any difference to the phototransport behavior of highly crystallized μc-Si:H films. However, the concentration of DB does significantly influence the behavior of phototransport properties, and the best results were obtained using $N_D$ = $10^{14}$ cm$^{-3}$ for *type-B* and *C* materials, and $N_D$ = $10^{15}$ cm$^{-3}$ for *type-A* material.

Now let us study the individual results of each type of material. In our *type-A* μc-Si:H material, the effective DOS profile is quite similar to a-Si:H, just as its properties elucidated in the experimental study were reminiscent of a-Si:H. The characteristic width of CBT in this material is narrow. Such a narrow width of CBT is seen in heavily doped n-type a-Si:H. The approximate distribution of CBT states in *type-A* material derived by applying the Rose model on the experimental data, when embodied with the correct numerical values derived from the simulation results, has yielded the accurate picture of CBT. This would not have been possible without the modeling study due to the lack of necessary information required to match the characteristic slopes of the tail states obtained for different energetic positions with some real experimental values of DOS at any position of the gap.

The above-mentioned valuable information regarding the CBT was elicited in *type-B* material as well. The corroboration of the experimental results by the simulated results in the *type-B* material validates our arguments regarding the VBT states, that is, the more numerous unpercolated paths present in this material give rise to two valence band tail slopes. The value of the DOS of the deeper VBT is lower in the film, and the steeper shallow tail states in *type-B* material plunging deeper as they fall from the band edge. A noteworthy point here is that the phototransport properties of this *type-B* μc-Si:H material are similar to what was observed by Balberg et al. for their μc-Si:H material. Moreover, the structural properties are also similar, in that, we have observed the absence of any distinguishable amorphous phase in our *type-B* μc-Si:H as well. However, we contend that the shapes of CBT and VBT used for model simulations of our material are different from those of Balberg et al. case.[22] The model simulation of Balberg et al. was based on the experimentally determined phototransport properties of majority carriers (obtained from SSPC) and as well as minority carriers (obtained from SSPG). In contrast, our simulation is based on the information obtained about majority carriers only. In the case of Balberg et al., though there are a larger number of holes than electrons, but SSPC data has been used to elicit information about the phototransport properties of majority carriers only, thus extracting information about one type of carrier only from a process that is an integrated behavior of both types of carriers participating in it. Therefore, it is essential to see if our model simulation [using the DOS shown in Fig.

11(a)] can facilitate the elicitation of phototransport properties of minority carriers as well. The Fig. 16(a) shows the temperature dependence of $(\mu\tau)_{hole}$ and $(\mu\tau)_{electron}$ obtained for the generation rate of $G_L = 10^{20}$ cm$^{-3}$s$^{-1}$. The behavior of temperature dependence of $(\mu\tau)_{hole}$ is found to be very close to that obtained experimentally as well as quantitatively (for the best case of Gaussian VBT with a width of 0.03 eV) by Balberg *et al.* A similar agreement is also observed between the results obtained for the temperature dependence of $(\mu\tau)_{electron}$ from our model simulation and the results so far shown by Balberg *et al.* In Fig. 16(b) we have presented the temperature dependence of $(\gamma)_{hole}$. The peculiarity in the $(\gamma)_{hole}$ in the case of Balberg *et al.* was that its value is less than 0.5 throughout the range of measurement temperature. Interestingly, the results obtained by us are very close to their observations even though our selections of DOS distributions are very different. In contrast to the *Gaussian* VBT that was considered by Balberg *et al.* to be the sole choice to reproduce the experimental features of phototransport properties in *type-B* µc-Si:H material, we have considered a structured VBT which has an exponential distribution in both the VBT regions. There is a strong ground for the selection of such a VBT as stated above, whereas the presence of Gaussian type VBT in µc-Si:H has not yet been experimentally confirmed.

Moreover, Balberg *et al.* considered a very low value of $\mu_{hole}$ (10$^{-4}$ cm$^2$/V-sec), which is rather beyond the scope of physical credibility for a reasonably good crystallized µc-Si:H. They attributed the choice of this low value of $\mu_{hole}$ to the newly found narrow Gaussian VBT and argued it to be a necessary constraint to obtain a better quantitative agreement with the experimental results. In contrast, we have chosen the value of $\mu_{hole}$ = 0.1 cm$^2$/V-sec and $\mu_{electron}$ = 5 cm$^2$/V-sec (obtained from TRMC experiment), which is frequently observed experimentally for µc-Si:H system. Therefore, at this juncture we would suggest that the effective DOS so far constructed to deduce the phototransport properties of *type-B* µc-Si:H can also be used as a DOS for a standard µc-Si:H, i.e., µc-Si:H material that is commonly studied in most of the laboratories worldwide, which has a microstructure similar to our *type-B* µc-Si:H. In addition, the useful information elicited about minority carriers from our model can be considered authentic, as it is similar to the information provided by Balberg *et al.* However, in this work we are not going to present the results of phototransport properties of minority carriers of other types of µc-Si:H due to the lack of sufficient experimental inputs.

Now let us discuss about the proposed effective DOS of fully crystallized *type-C* µc-Si:H material, the bulk of which contains a large number of densely packed large crystalline columns with a reduced amount of defects. According to the Fig. 14(a), the DOS profile consists of a very steep CBT, which is followed by a deeper CBT having lower density of states, with a slowly decaying slope.

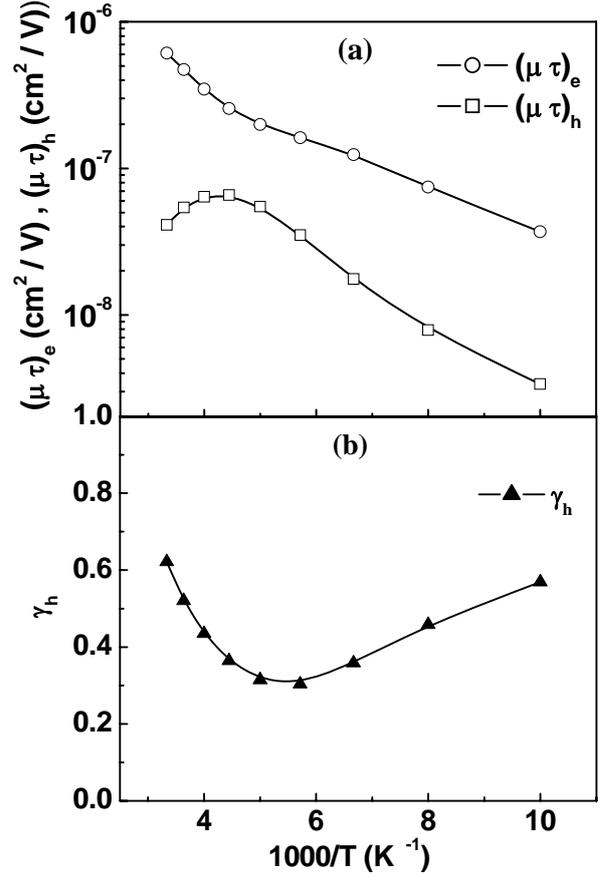

FIG. 16. Simulated phototransport properties of *type-B* µc-Si:H material using the DOS as shown in Fig. 11(a) (Case-B2). The part (a) simulated temperature dependent $(\mu\tau)_{hole}$ and $(\mu\tau)_{electron}$ for the light intensities $G_L = 10^{20}$ cm$^{-3}$s$^{-1}$ and (b) simulated temperature dependent light intensity exponent of *minority carrier* ($\gamma_h$).

Qualitatively we had asserted that in fully crystallized *type-C* µc-Si:H, higher density of available free carriers and low value of defect density can create the possibility of such a steep CBT (§III.D.3). This can be substantiated and understood in detail only when we see the carrier density profile, which will be presented later in this section. Regarding the VBT region of our proposed DOS of *type-C* material, it is seen as a hybrid form having two distinct parts with different slopes; one with a sharper slope near the edge and another with a less steep slope at deeper energy level. We have proposed a similar hybrid form of VBT structure for *type-B* material as well, but the difference is that the deeper VBT has a higher DOS in *type-C* material than that in *type-B* material. Analytically, we had suggested that this difference is due to the *type-C* material having acquired a larger number of percolated paths than *type-B* material (§III.D.3). Therefore, we can conclude here that the proposed effective DOS [Fig. 14(a)] for this highly crystallized densely packed *type-C* µc-Si:H material (which is often considered comparable to the so called hydrogenated poly-silicon), not only has a strong analytical support for its physical plausibility, but also is capable of producing simulated phototransport properties that corroborate with the experimental results.

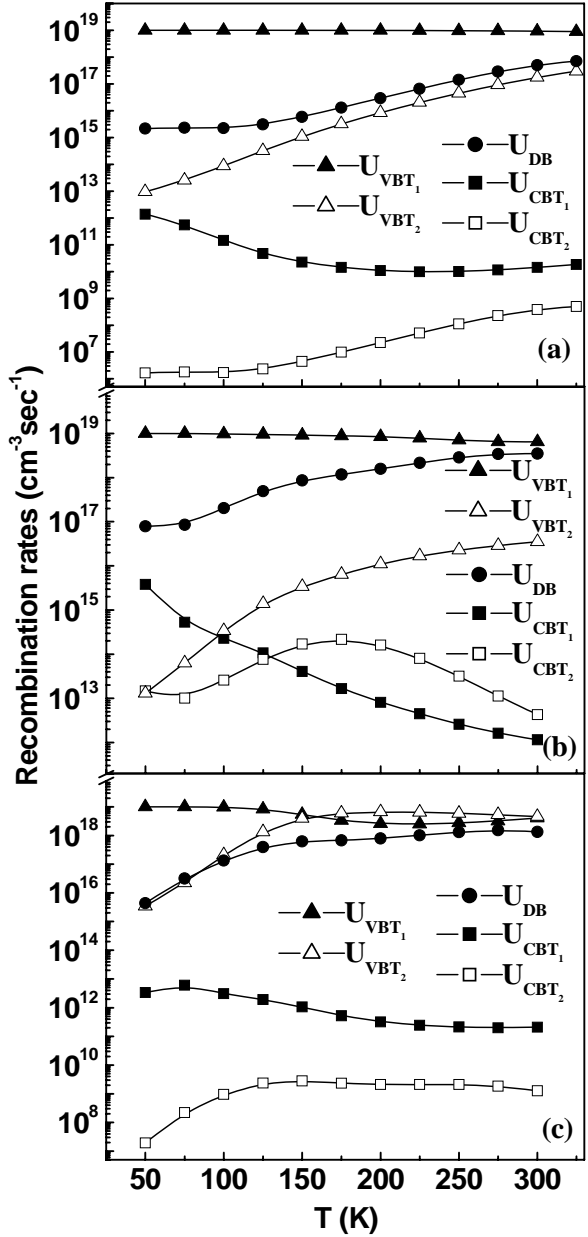
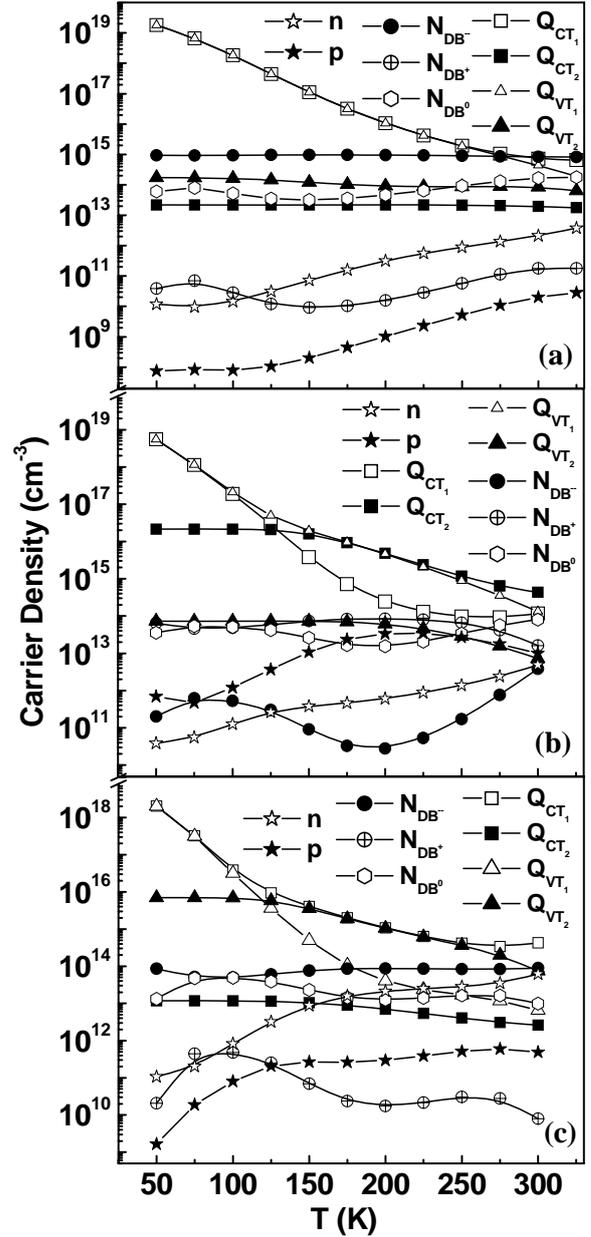

FIG. 17. Simulated recombination rates of (a) *type-A* μc-Si:H (b) *type-B* μc-Si:H and (c) *type-C* μc-Si:H. The generation rate $G_L = 10^{19}$ cm$^{-3}$s$^{-1}$.

FIG. 18. Simulated carrier densities of (a) *type-A* μc-Si:H (b) *type-B* μc-Si:H and (c) *type-C* μc-Si:H. The generation rate $G_L = 10^{19}$ cm$^{-3}$s$^{-1}$.

So far we have had a detailed discussion regarding the choice and suitability of the constructed DOS profiles of our μc-Si:H system. Now it would be highly interesting to see how the recombination processes actually take place in the various parts of DOS in different μc-Si:H systems. The simulated temperature dependence of re-combination rates of *types-A, B* and *C* of μc-Si:H material for a particular generation rate, $G_L = 10^{19}$ cm$^{-3}$s$^{-1}$ are shown in Fig. 17(a), (b) and (c) respectively. These recombination rates were deduced from the model simulation using the respective DOS models of each type. The temperature dependence of recombination rates of *type-A* μc-Si:H shown in Fig. 17(a) reveal that dominant recombination takes place in shallow VBT states and it remains constant in the whole range of temperature under study. On the other hand, the recombination rate of DBs starts rising with the increase in temperature. Similar temperature dependent behavior of recombination rate is also shown by deeper VBT states and it can be seen that above 150K the recombination rate of these states are very close to that of DBs. However, no such cross-over between the recombination rates of any of the localized states is seen near the onset of TQ, which is usually, but not always, seen in *a*-Si:H. This will be further discussed when we come to the carrier density profile of this case. The recombination mechanism of *type-B* μc-Si:H material also shows the dominant recombination taking place in the shallower VBT states, as seen in Fig. 17(b). In contrast, the temperature dependence of recombination rates of *type-C* μc-Si:H material presented in Fig. 17(c) shows that at low temperatures the recombination is mostly dominated by the shallower valence band tail states, but as the temperature increases, the deeper VBT states starts playing a role in the recombination process. Later, at higher temperatures, the recombination rate in deeper VBT states becomes higher than that in shallower VBT states. This is the temperature from where the $\sigma_{ph}(T)$ for this particular generation rate starts changing its trend, its monotonic increase gradually slowing down, which eventually at a still higher temperature, culminates in *TQ* behavior. Thus, it suggests that the *TQ* behavior in this type of material occurs when there is a transfer of the recombination traffic from shallower VBT states to deeper VBT.

At this point, it is evident from the simulated temperature dependences of the recombination rates that though both the *types A* and *C* of μc-Si:H material evince TQ behavior, the cause is different for each case. Now let us look at the temperature dependence of the simulated carrier densities of *types A, B* and *C* μc-Si:H respectively in Figs. 18 (a), (b) and (c). As we have seen in the DOS profiles of these types of material, the CBT of *type-C* is very steep compared to either of the other two types. In both the cases (*type-A* and *C*), the width of CBT is narrower than that of VBT. Its effect can be seen in the carrier density graphs of these two cases [Fig. 18 (a) and (c)], where most of the DBs are seen to be negatively charged as more electrons are captured by DBs than holes. Consequently, it causes a large increase in $N_{DB}^{-}$ together with a decrease in $N_{DB}^{+}$ states in the gap, which in turn shows a lower DOS near the CB edge leading to steeper CBT in these two types of μc-Si:H material. In *type-A* μc-Si:H, the trapped carrier densities in shallower CBT ($Q_{CT1}$) and shallower VBT ($Q_{VT1}$) are almost the same till *T*= 270K, after which they part ways, and simultaneously, $N_{DB}^{-}$ starts equaling $Q_{CT1}$. This is the temperature where we have observed TQ in simulated $\sigma_{ph}(T)$ of *type-A* μc-Si:H. Whereas in *type-C* μc-Si:H, the equality in $Q_{CT1}$ and $Q_{VT1}$ is broken when the carrier densities of deeper VBT ($Q_{VT2}$) start becoming equal in number to $Q_{CT1}$. The temperature, at which this equality is achieved, is the same where the onset of TQ is seen in *type-C* μc-Si:H. These $Q_{VT2}$ are basically the safe hole traps which do not play a role in recombination at low temperatures, but at higher temperatures they provide a dominant recombination path in the phototransport process.

This quantitative information derived from the above analyses affirm the validity of the effective DOS profiles and the recombination processes occurring therein, which we have proposed for all the three microstructurally different μc-Si:H materials.

## V. SUMMARY AND CONCLUSION

The steady state photoconductivity as a function of temperature and light intensity was measured on plasma deposited highly crystalline undoped μc-Si:H samples prepared under a wide array of deposition conditions to yield different types of film microstructure. We employed a variety of characterization tools to probe the film microstructure at different length scales. We classified the dominant features of film microstructure with correlative dark electrical transport properties, into three categories representing the whole range of μc-Si:H films, designated as *types A, B* and *C* in this study. We studied the optoelectronic properties of these well characterized μc-Si:H films having varying degrees of crystallinity and tried to identify the role of microstructure in determining the electron transport behavior using dark conductivity and photoconductivity as functions of several discerning parameters such as temperature, wavelength and intensity of probing light. In order to understand the experimental results, we constructed a computer simulation model using Shockley-Read statistics in steady state conditions to determine the recombination processes.

Different phototransport behaviors were observed in films belonging to different types of μc-Si:H material. In *type-A* μc-Si:H films, light intensity exponent (γ) lies between 0.5 and 1, and $\sigma_{ph}(T)$ shows thermal quenching (*TQ*) effect. In *type-B* μc-Si:H material, 0.5 < γ < 1 with *No TQ* effect; and in *type-C* μc-Si:H material, 0.15 < γ < 1 with a *TQ* effect, which is anomalous, are observed. We have presented cogent evidence that the effective DOS profiles are different for each microstructurally different type of μc-Si:H material. The light intensity exponent (γ) derived from photogeneration-rate dependence of the photocurrent describes the power law behavior in μc-Si:H, and its study elucidates the physical properties of this material. The Rose model developed for exponentially distributed gap-states was found to be applicable in *type-A* and *type-B* μc-Si:H system. We have experimentally obtained the DOS profiles of *type-A* and *B* material below the conduction band edge using the Rose model. However, in *type-C* μc-Si:H material, the Rose model is not found to be valid as $kT_c < kT$, which implies a very steep CBT ($kT_c < 0.02$ eV). To explain the experimental findings, we have proposed structured valence band tails having distinct shallow and deeper states in *type-B* and *type-C* materials, and a wider VBT, similar to that found in *a*-Si:H in *type-A* material. We have been able to construct complete effective DOS distributions of all the three types of $\mu$c-Si:H systems. Our proposed effective DOS of $\mu$c-Si:H system exhibits structured valence band tails; a sharper, shallow tail originating from grain boundary defects and another less steeper, deep tail associated with the defects in the columnar boundary regions. Depending on the film microstructure and possible percolation paths available in the particular system, the DOS values of these shallow and deeper VBT may differ.

The results of our model simulation of photoconductivity were found to be in excellent agreement with the experimental findings. The results of simulated phototransport behavior of *type-B* and *C* $\mu$c-Si:H materials validate our contention that in the highly crystalline $\mu$c-Si:H material, it is the distribution of the localized states in the gap, rather than the gap width itself, that has a bearing on the recombination processes, and thereby on the phototransport properties. The defect states located within the disordered phase of the material affect the recombination processes, wherever the disorder phase exists within the material (e.g., columnar or grain boundaries). Our model simulation study demonstrated that a change in the DB distribution ($W = 0.3$ eV vs $W = 0.15$ eV) has little influence on the phototransport behavior of highly crystallized $\mu$c-Si:H films, though the concentration of DB does have an important effect. We have contended that the underlying causes of $TQ$ are different for *type-A* and *type-C* $\mu$c-Si:H material. The simulated temperature dependences of recombination rates clearly show that in *type-A* material, the dominant recombination takes place in shallow VBT states and remains constant for the whole range of temperature under study. In contrast, in *type-C* material, recombination is mostly dominated by the shallower VBT states at lower temperatures, but the deeper VBT states start playing a role in the recombination process as the temperature increases. The trapped carrier densities in these deeper VBT states act as safe hole traps, which do not play a role in recombination at low temperatures, but govern the recombination process at higher temperatures. This elucidates well the cause of anomalous behavior experimentally observed in phototransport properties of all the samples of the fully crystallized *type-C* $\mu$c-Si:H material.

The DOS profile of *type-B* $\mu$c-Si:H proposed by us can be used for any standard plasma deposited $\mu$c-Si:H having the microstructure commonly possessed by such films, which is typified by the *type-B* $\mu$c-Si:H of this study. Our simulation model can provide reliable information about the phototransport properties of minority carriers as well.